%% file: main-11.tex
\documentclass[aps,prd,reprint,superscriptaddress,nofootinbib,longbibliography]{revtex4-2} 

\usepackage{mlmodern}
\usepackage{lipsum}
\usepackage{xcolor}
\usepackage{mathrsfs}
\usepackage{booktabs}
\usepackage{subcaption}
\usepackage{ulem}
\input{preamble}
\usepackage{footnotehyper}

\makeatletter

\newcommand{\linkedfootnotemark}[1]{%
  \stepcounter{footnote}%
  \protected@xdef\@thefnmark{\thefootnote}%
  \expandafter\xdef\csname linkedfn@number@#1\endcsname{%
    \number\value{footnote}%
  }%
  \leavevmode
  \ifhmode
    \edef\@x@sf{\the\spacefactor}\nobreak
  \fi
  \hyperlink{linkedfn:#1}{\@makefnmark}%
  \ifhmode\spacefactor\@x@sf\fi
  \relax
}

\newcommand{\linkedfootnotetext}[2]{%
  \footnotetext[\csname linkedfn@number@#1\endcsname]{%
    \hypertarget{linkedfn:#1}{}%
    #2%
  }%
}

\makeatother

\usepackage{caption}
\begin{document}

\makeatletter
\long\def\@makecaption#1#2{%
    \par
    \vskip\abovecaptionskip
    \begingroup
        \small\rmfamily
        \samepage
        \flushing
        \parindent\z@
        \let\footnote\@footnotemark@gobble
        \@make@capt@title{#1}{#2}\par
    \endgroup
    \vskip\belowcaptionskip
}
\makeatother

\title{Rolling Galileons: Evolving Braiding Strength for Viable Dark Energy}


\author{James Hallam}
\email{james.hallam@port.ac.uk}
\affiliation{Institute of Cosmology and Gravitation, University of Portsmouth, Portsmouth PO1 3FX, U.K.}

\author{Krishna Naidoo}
\affiliation{Institute of Cosmology and Gravitation, University of Portsmouth, Portsmouth PO1 3FX, U.K.}
\affiliation{Max-Planck-Institut f\"ur Astronomie, K\"onigstuhl 17, 69117 Heidelberg, Germany}

\author{Sergi Sirera}
\affiliation{Institute of Cosmology and Gravitation, University of Portsmouth, Portsmouth PO1 3FX, U.K.}

\author{Tessa Baker}
\affiliation{Institute of Cosmology and Gravitation, University of Portsmouth, Portsmouth PO1 3FX, U.K.}

\date{\today}

\begin{abstract}
Motivated by growing observational indications that dark energy may be dynamical, we introduce Rolling Galileon gravity: a minimal shift-symmetry-breaking extension of the cubic Galileon in which the coupling coefficients are allowed to vary, giving rise to an evolving braiding strength. The full theory space, shown to be closed under field redefinitions, is characterised by two functions. We derive analytical conditions to satisfy three phenomenological requirements: i) a phantom-crossing equation of state at late times, ii) a positive integrated Sachs-Wolfe signature, and iii) absence of pathologies in the screened scalar force in cosmic voids. We show that these conditions are collectively satisfied by an increasing braiding strength relative to the kinetic sector. A Bayesian analysis of minimal Rolling Galileon models finds that they can satisfy the viability requirements i)-iii) whilst providing an acceptable fit to expansion-history data.
\end{abstract}

\maketitle

\newpage

\section{Introduction}
\label{sec:intro}

Recent cosmological observations re-emphasise the question of whether late-time cosmic acceleration is truly described by a cosmological constant, or instead by dynamical dark energy. Whilst the standard $\Lambda$CDM model remains remarkably successful across a broad range of data, current combinations of baryon acoustic oscillations (BAO), supernova, and cosmic microwave background (CMB) measurements have increasingly been interpreted as favouring an equation of state that evolves with cosmic time, with indications of a late-time crossing of the phantom divide~\cite{Calderon_2024, DESI:2025zgx, DES_2026}.\linkedfootnotemark{phantom} 
Such indications thus motivate the search for a microphysical origin of this behaviour, see e.g.~\cite{Cehn_2025, Goh:2025quintom, Hallam2025, Nojiri_2025, Yang_2025, Guedezounme_2026, HiCOLA:Krishna.paper}. 

Horndeski gravity is the most general four-dimensional scalar-tensor theory yielding second-order field equations~\cite{Horndeski:1974wa}. In this work, we restrict ourselves to its luminal subclass, in which tensor modes propagate at the speed of light~\cite{Abbott_2017, Baker:2017hug, Creminelli:2017sry}.\linkedfootnotemark{gw} The luminal Horndeski action may be written schematically as
\begin{align}
    S = \int d^4x\sqrt{-g}\Big[&G_4(\phi)R+K(\phi, X) 
    \nonumber\\
    & \qquad
    -G_3(\phi, X)\Box\phi\Big] +S_{\rm m}, 
    \label{eq:luminal_Horndeski}
\end{align}
where $g_{\mu\nu}$ is the spacetime metric, $g=\det (g_{\mu\nu})$, $R$ is its Ricci scalar, $S_{\rm m}$ is the matter action, and $\phi$ is the scalar field. Its kinetic term and d'Alembertian are defined respectively by $X\coloneqq  -\frac12 g^{\mu\nu}\partial_\mu\phi\partial_\nu\phi$ and $\Box \phi \coloneqq g^{\mu\nu} \nabla_\mu \nabla_\nu \phi$.
The simplest extension to General Relativity (GR) is pure $k$-essence, for which $G_3=0$ and 
$G_4=\text{const}$,
so the scalar evolves only through $K(\phi,X)$. Although this class is very flexible at the level of the background expansion~\cite{Armendariz-Picon:2000nqq}, it cannot realise a stable phantom crossing~\cite{Vikman_2005}. One must therefore either make $G_4$ field dependent, corresponding to non-minimal coupling~\cite{Brans:1961sx, Perrotta:1999am, Perivolaropoulos:2005yv}, or activate $G_3$, corresponding to kinetic gravity braiding~\cite{C_dric_Deffayet_2010, Pujolas:2011he}.

Non-minimally coupled models with $G_4(\phi)$ can successfully accommodate current expansion-history data~\cite{Ye:2024ywg, Wolf:2024stt, Wolf:2024eph, Gu:2025xie}. However, as a result of the associated scalar field coupling to gravity, the growth of structure is also modified, and joint fits of distances and growth warrant caution~\cite{Linder:2025zxb, Garcia-Garcia:2026nzy}.
An alternative is to remain minimally coupled ($G_4=\text{const}$) and instead use the braiding sector $G_3$. In this subclass, 
stable phantom crossing histories consistent with background, growth, and integrated Sachs-Wolfe (ISW) data are known to exist~\cite{Cataneo:2025vae}, with Asymptotically Cubic Galileons (ACG) providing a first Lagrangian realisation~\cite{HiCOLA:Krishna.paper}. 
Table~\ref{tab:model_comparison} lists candidate models of each kind, alongside comments on whether they admit a stable phantom crossing compatible with other observations. 
Overall, these results identify minimally coupled braiding as a promising subclass for attaining a healthy phantom crossing.

\definecolor{forestgreen}{RGB}{34,139,34}
\definecolor{darkorange}{RGB}{245,145,0}
\definecolor{darkred}{RGB}{139,0,0}

\newcommand{\good}[1]{\textcolor{forestgreen}{#1}}
\newcommand{\caution}[1]{\textcolor{darkorange}{#1}}
\newcommand{\bad}[1]{\textcolor{darkred}{#1}}

\newcommand{\cmark}{\good{\ensuremath{\checkmark}}}
\newcommand{\xmark}{\bad{\ensuremath{\times}}}
\newcommand{\pmark}{\ensuremath{\sim}}

\begin{table*}[t!]
    \centering
    \setlength{\tabcolsep}{4pt}
    \renewcommand{\arraystretch}{1.08}
    \footnotesize

    \begin{tabular}{p{4.5cm} c c c c p{6.6cm}}
        \hline\hline
        Model & $K$ & $G_3$ & $G_4$ & PC & Comment \\
        \hline
        $k$-essence~\cite{Vikman_2005}
          & $K(\phi,X)$ & $0$ & $M_{\rm Pl}^2/2$ & \xmark & \bad{Cannot yield a stable crossing.} \\
        Non-minimal coupling~\cite{Ye:2024ywg, Wolf:2024stt, Wolf:2025jed, Wolf:2024eph, Linder:2025zxb, Garcia-Garcia:2026nzy, Gu:2025xie}
          & $K(\phi,X)$ & $0$ & $G_4(\phi)$ & \cmark & \bad{Growth modified;} \caution{requires model-dependent care.} \\
        Cubic Galileon~\cite{Nicolis_2009, Deffayet:2009wt, HiCOLA:Krishna.paper}
          & $-k_0 X$ & $X$ & $M_{\rm Pl}^2/2$ & \xmark & \bad{No crossing; viable fits prefer $\Lambda$CDM limit.} \\
        Shift-symmetric braiding~\cite{Traykova_2021, Linder:2025pqt}
          & $K(X)$ & $G_3(X)$ & $M_{\rm Pl}^2/2$ & \xmark & \bad{Generically no crossing whilst shift-symmetric.} \\
        Massive cubic Galileon~\cite{Garcia-Garcia:2026nzy, Wolf_2026} 
          & $-k_0X-1/2m^2\phi^2$ & $X$ & $M_{\rm Pl}^2/2$ & \cmark & \bad{Growth modified;} \caution{void-pathology not assessed.} \\
        Asymptotically Cubic Galileon~\cite{HiCOLA:Krishna.paper}
          & $-k_0\mathcal K(\phi)X$ & $\mathcal G(\phi)X$ & $M_{\rm Pl}^2/2$ & \cmark & \good{Positive ISW; minor growth effects;} \bad{void-pathology.} \\
        Rolling Galileons~[this work]
          & $-k(\phi)X+q(\phi)X^2$ & $X$ & $M_{\rm Pl}^2/2$ & \cmark & \good{Positive ISW; able to remain healthy in voids.$^\dagger$} \\[0.2em]
        \hline\hline
        \multicolumn{6}{l}{\footnotesize $^{\dagger}$ In this work, a particular Rolling Galileon model is found to satisfy all viability conditions; see Table~\ref{tab:RG_results}.} \\
    \end{tabular}
    
    \caption{\label{tab:model_comparison}    
    Luminal Horndeski models, in the notation of~\eqref{eq:luminal_Horndeski}, confronted with a stable phantom crossing (PC), together with comments on their potential viability. Symbols: \cmark\ achieved and consistent with current data; \xmark\ structurally obstructed.}
\end{table*}

\linkedfootnotetext{phantom}{The value $w_\mathrm{DE} = -1$ of the dark energy equation of state separates the quintessence $(w_\mathrm{DE}>-1)$ and phantom $(w_\mathrm{DE}<-1)$ regimes. A \textit{phantom crossing} is simply the transition across it.}

\linkedfootnotetext{gw}{Although the GW170817 bound was obtained at frequencies near the expected cutoff of dark energy effective field theories, making its extrapolation to cosmological scales rather non-trivial~\cite{deRham:2018red}, we assume the luminal restriction as a conservative choice. This avoids the reliance on an unspecified ultraviolet completion. For potential observational tests on this assumption, see~\cite{Harry:2022zey,Baker:2022rhh,Baker:2022eiz,Sirera:2023pbs,Atkins:2024nvl,Kobayashi:2025evr}.}


This work constructs the minimal yet complete
theory space 
obtained by breaking the 
shift symmetry of the simplest kinetic-braiding structure.
It contains ACG as a special case, and isolates the components responsible for viable phantom crossings.
The simplest kinetic-braiding theory is the cubic Galileon~\cite{Nicolis_2009, Deffayet:2009wt, Appleby:2011aa}. This model is particularly appealing because the Galileon symmetry protects its derivative interactions from large quantum corrections~\cite{Goon_2016}, whilst its tracker solution fixes its single free parameter~\cite{Wright:2022krq}.
Its effects in dense environments are also screened via the Vainshtein mechanism ~\cite{Vainshtein:1972sx, Babichev:2013usa, Sirera:2026klo}.
Moreover, it constitutes the only theory known to admit stable black hole solutions that can be parametrically connected to cosmological regimes~\cite{Smulders:2026qwc,Smulders:2026bya}.
Nevertheless, it remains eternally phantom~\cite{Traykova_2021}. 
Even with non-standard kinetic structure, a phantom crossing stays obstructed whilst the shift symmetry is intact~\cite{Linder:2025pqt}; crossing therefore requires breaking it. Breaking shift symmetry through the addition of a scalar potential achieves a stable crossing, but modifies the growth of structure~\cite{Tsujikawa:2025wca, Wolf_2026}. In this work, we instead break shift symmetry by promoting the constant coupling coefficients of the cubic Galileon model to functions of the scalar field. We call the resulting theory space \textit{Rolling Galileon} gravity (RG).
We will see that a single rolling trend suffices to drive a stable phantom crossing whilst maintaining a positive ISW signature and a healthy screened force in voids, with a good fit to observational data.

This paper is organised as follows. 
In Section~\ref{sec:ACG}, we characterise the full Rolling Galileon theory space with two invariant functions.
In Section~\ref{sec:phantom_crossing}, we derive analytical conditions on these functions from phantom crossing, ISW, and void viability requirements, and identify their mutual compatibility. 
In Section~\ref{sec:carpets}, we numerically explore the resulting parameter space of a simple rolling model via MCMC.
We conclude in Section~\ref{sec:conclusions} with a summary of the key findings and future directions.

In this paper, we adopt the mostly-plus metric signature $(-+++)$, work in natural units with $c=1$, and define the reduced Planck mass as $M_\text{Pl}^2~\coloneqq(8\pi G)^{-1 }$.

\section{Rolling Galileon Gravity}
\label{sec:ACG}

Within the scalar sector, the shift-symmetric cubic Galileon constitutes a genuine one-parameter theory, before imposing any tracker condition. 
Allowing its constant coupling coefficients to roll with the scalar leads to a self-contained extension of this theory, described by two independent functions. In this Section, we identify the scalar-redefinition invariant quantities that fully parametrise this theory space, and use them to characterise physical departures from the cubic Galileon. We start by reviewing the shift-symmetric theory itself.


\subsection{Reviewing the Cubic Galileon}
Allowing also for a cosmological constant $\Lambda$, the scalar contribution $\mathcal L_\varphi$ enters the full theory via
\begin{align}
    S
    =
    \int d^4x \sqrt{-g}
    \left[
        \frac{M_\text{Pl}^2}2 R
        - \Lambda + \mathcal L_\varphi
    \right]
    + S_{\rm m},
    \label{eq:ACG_full_action}
\end{align}%
where $S_{\rm m}[g_{\mu\nu},\psi]$ denotes the matter action for matter fields $\psi$, assumed to be minimally coupled to the spacetime metric. This restriction is well motivated, as coupled Galileons are tightly constrained~\cite{Brax:2014vla}. 

Throughout this section, we denote by $\varphi$ an arbitrary scalar field, and reserve $\phi$ for the redefined field. We correspondingly define the kinetic terms
\begin{equation}
    \mathcal X\coloneqq -\frac12 g^{\mu\nu}\partial_\mu \varphi\partial_\nu \varphi,
    \qquad
    X\coloneqq -\frac12 g^{\mu\nu}\partial_\mu \phi\partial_\nu \phi.
\end{equation}
Furthermore, subscripts of a field denote derivatives with respect to that field, e.g. $F_\varphi = \partial F/\partial\varphi$ with $F(\varphi)$ an arbitrary function of the field.

The cubic Galileon considered here may be written 
as
\begin{equation}
    \mathcal L_\varphi
    = -a \mathcal X - b \mathcal X \Box\varphi,
    \label{eq:general_CG}
\end{equation}
where $a$ and $b$ are positive constants. This apparent two-parameter freedom is a result of the choice of scalar field parametrisation. Under a linear field redefinition, $\varphi= \varphi(\phi)$ with $\varphi_\phi=\text{const}$ and $\varphi_{\phi\phi}=0$,
one has
\begin{equation}
    \mathcal X = \varphi_\phi^2 X,
    \qquad
    \Box\varphi = \varphi_\phi \Box\phi ,
\end{equation}
and hence
\begin{equation}
    \mathcal L_\phi
    = -a \varphi_\phi^2X - b \varphi_\phi^3X \Box\phi.
\end{equation}
Choosing $\varphi_\phi = b^{-1/3}$
normalises the braiding term, giving
\begin{align}
    \mathcal L_\phi = - k_0 X - X \Box\phi,
    \qquad
    k_0 \coloneqq \frac{a}{b^{2/3}}.
    \label{eq:k-t-s}
\end{align}
Thus, modulo constant scalar field rescaling, the cubic Galileon scalar sector is characterised by the constant \textit{kinetic-to-braiding} strength $k_0$. In cosmological applications this parameter may be further fixed by imposing the tracker solution~\cite{Wright:2022krq, HiCOLA:Krishna.paper}, but we do not assume that restriction here.

\subsection{The General Closed Theory Space} 
\label{subsec:closed_theory_space}
The linear field redefinition is only the simplest instance of scalar field reparametrisation freedom. 
Allowing the coupling coefficients to depend explicitly on the scalar, we now consider a general field redefinition $\varphi=\varphi(\phi)$, for which $\varphi_{\phi\phi}$ need not vanish. 
Such nonlinear redefinitions generate an additional quadratic-kinetic $X^2$ contribution originating from the braiding term.
In the interest of brevity, we hereafter refer to $X^2$ as simply the \textit{quadratic} term. Hence, let us consider the general Lagrangian
\begin{equation}
    \mathcal L_\varphi
    =-A(\varphi) \mathcal X - B(\varphi) \mathcal X  \Box\varphi + C(\varphi) \mathcal X^2,
    \label{eq:general_ACG}
\end{equation}
where $A$, $B$, and $C$ are arbitrary functions of $\varphi$. Under this generic field redefinition one has
\begin{align}
    \mathcal X &=  \varphi_\phi^2 X, 
    \label{eq:X_transform}
    \\
    \Box \varphi &=  \varphi_\phi \Box \phi - 2  \varphi_{\phi\phi}X.
    \label{eq:boxX_transform}
\end{align}
Substituting~\eqref{eq:X_transform}-\eqref{eq:boxX_transform} into~\eqref{eq:general_ACG}, the Lagrangian retains the same operator form,
\begin{equation}
    \mathcal L_\phi
    =-\widetilde A(\phi) X - \widetilde B(\phi) X \Box\phi + \widetilde C(\phi) X^2,
    \label{eq:general_transformed_ACG}
\end{equation}
with the transformed coefficient functions
\begin{align}
    \widetilde A &= A \varphi_\phi^2, 
    \label{eq:A_transformation_law}
    \\
    \widetilde B &= B \varphi_\phi^3, 
    \label{eq:B_transformation_law}
    \\
    \widetilde C  
    &= C \varphi_\phi^4 + 2B \varphi_\phi^2 \varphi_{\phi\phi}.
    \label{eq:C_transformation_law}
\end{align}
We therefore see that the Lagrangian structure~\eqref{eq:general_transformed_ACG}, with the inclusion of the $X^2$ operator, is closed under field redefinitions. This is therefore the complete operator basis at the lowest polynomial order containing braiding. Allowing an independent arbitrary function of $\varphi$ as the coefficient of each operator then yields the most general theory within this closed truncation. Of these three coefficient functions, one is redundant under field redefinitions, leaving two independent functions whose invariant combinations we now identify.

\subsection{A Braiding-Normalised Parametrisation} 
\label{subsec:invariant_functionals}

We now use the braiding strength $B$ to fix the scalar-field normalisation and construct a pair of field-redefinition invariant functions. The choice here is not unique, but is natural here because Rolling Galileons are organised as departures from the cubic Galileon within the braiding sector. The resulting functions measure the remaining kinetic strengths relative to braiding and provide one complete invariant parametrisation of the closed two-function theory space. Here and throughout, “strength” refers to the coupling coefficients, rather than to the relative magnitudes of the operators evaluated on a particular cosmological solution, which additionally depend on $X$; for a complementary background-level decomposition into contributions, see Ref.~\cite{Calderon:2026hbr}.

\subsubsection{Generalised Kinetic-to-Braiding Strength}
We begin with the generalisation of the cubic Galileon kinetic-to-braiding constant. Let us momentarily suppress the quadratic term and consider only the generalised cubic Galileon operators. Using~\eqref{eq:A_transformation_law} and~\eqref{eq:B_transformation_law}, we have
\begin{align}
    \mathcal L_\phi
    &\supset-\widetilde A X - \widetilde B X \Box\phi, \nonumber\\
    &=- A \varphi_\phi^2 X - B\varphi_\phi^3 X \Box\phi,
    \label{eq:generalised_cubic_Galileon_transform}
\end{align}
where recall that the dependencies for tilded and untilded coefficients are respectively $\widetilde A(\phi)$ and $A(\varphi(\phi))$.
In direct analogy with the cubic Galileon normalisation $\varphi_\phi = b^{-1/3}$ (see Eq.~\eqref{eq:k-t-s}),
we may choose
\begin{equation}
    \varphi_\phi=B^{-1/3}.
    \label{eq:braiding_coordinate}
\end{equation}
We are essentially using the braiding coefficient as the normalising scale, and thus assuming that the braiding sector is always active, $B(\varphi)\neq0$. The two signs of $B$ therefore define disconnected patches of the field space. In this work, we choose the $B>0$ patch, which reduces to the $b>0$ convention in the constant-coefficient theory~\eqref{eq:general_CG}. Indeed, the $B<0$ patch can be treated analogously, but a smooth crossing between them is prohibited in this choice of coordinate, for a crossing between them would pass through $B=0$ where the braiding coordinate becomes singular. This is not a physical pathology; it only means this is no longer a good coordinate choice there.

In this braiding coordinate~\eqref{eq:braiding_coordinate},
\begin{align}
    \mathcal L_\phi
    &\supset- \tilde k X - X \Box\phi,
    \qquad
    \tilde k\coloneqq\frac{\widetilde A}{\widetilde B^{2/3}}.
    \label{eq:k_inv}
\end{align}
This is the generalisation of the cubic Galileon kinetic-to-braiding strength in~\eqref{eq:k-t-s}, now promoted from a constant to a field-dependent function. Its invariance follows as
\begin{equation}
    \widetilde k
    = \frac{\widetilde A}{\widetilde B^{2/3}} 
    = \frac{A \varphi_\phi^2}{B^{2/3}(\varphi_\phi^3)^{2/3}} 
    = \frac{A}{B^{2/3}} 
    = k.
    \label{eq:first_invariant}
\end{equation}
Thus, since $\tilde k (\phi) = k(\varphi(\phi))$, it provides a field-redefinition invariant characterisation of the kinetic-to-braiding strength throughout the closed theory space.

\subsubsection{Quadratic-to-Braiding Strength}
We now construct a second invariant to complete the characterisation of the closed theory space. This is associated with the new quadratic operator $X^2$, whose coefficient transforms according to~\eqref{eq:C_transformation_law}, 
\begin{align}
    \mathcal L_\phi
    &\supset \widetilde C X^2, \nonumber\\
    &= \left[C \varphi_\phi^4 
    + 2B \varphi_\phi^2 \varphi_{\phi\phi}\right] X^2.
\end{align}
Thus, the bare coefficient does not transform homogeneously. Part of the rolling braiding contribution is transferred into the coefficient of $X^2$.


Since $C$ does not transform homogenously, it cannot be normalised directly by a power of $B$ to form an invariant. We therefore seek a combination of $C$ and $B_\varphi$ that transform homogeneously.
Differentiating the transformed braiding coefficient $\widetilde B$ gives
\begin{align}
    \widetilde B_\phi
    &= B_\varphi\varphi_\phi^4
    + 3B\varphi_\phi^2\varphi_{\phi\phi}.
\end{align}
The inhomogeneous term in $\widetilde C$, proportional to $B\varphi_\phi^2\varphi_{\phi\phi}$, appears also in $\widetilde B_\phi$. Subtracting an appropriate combination therefore cancels it exactly,
\begin{align}
    \widetilde D
    &\coloneqq \widetilde C - \frac23 \widetilde B_\phi 
    = \left(
        C
        - \frac23 B_\varphi
    \right)\varphi_\phi^4
    = D\varphi_\phi^4.
\end{align}
The inhomogeneous terms cancel and $\widetilde D$ transforms homogeneously as $\varphi_\phi^4$.

We may therefore form the second invariant by normalising it by the braiding scale as
\begin{equation}
    \widetilde q
    \coloneqq
    \frac{\widetilde D}{\widetilde B^{4/3}}
    =\frac{D\varphi_\phi^4}{(B\varphi_\phi^3)^{4/3}}
    =\frac{D}{B^{4/3}}
    = q,
    \label{eq:second_invariant}
\end{equation}
which measures the coefficient multiplying $X^2$ with respect to braiding, thus defining the \textit{quadratic-to-braiding} strength.


The invariant pair $\{k(\phi),q(\phi)\}$ defined in~\eqref{eq:first_invariant} and~\eqref{eq:second_invariant} therefore provides a complete parametrisation of the closed two-function theory space on any patch with $B\neq0$. This theory space shall henceforth be known as \textit{Rolling Galileon} gravity (RG). It is the minimal closed shift-symmetry breaking extension of the cubic Galileon within the Kinetic Gravity Braiding (KGB) class~\cite{C_dric_Deffayet_2010, Pujolas:2011he}, or equivalently minimally coupled luminal Horndeski. 


\subsubsection{The Cubic Galileon Point and its Asymptotic Subclass}
Within the Rolling Galileon theory space, the shift-symmetric cubic Galileon is the special point at which the kinetic-to-braiding strength is constant and the quadratic-to-braiding strength vanishes:
\begin{equation}
    k(\phi) = k_0,
    \qquad
    q(\phi) = 0,
    \label{eq:CG_invariant_point}
\end{equation}
where $k_0$ is the same constant appearing in~\eqref{eq:k-t-s}.

A subclass of Rolling Galileons is obtained by demanding that the theory asymptote to this cubic Galileon point at early times. This defines the \textit{Asymptotically Cubic Galileon} (ACG) models studied observationally in the companion paper~\cite{HiCOLA:Krishna.paper}. Taking $0<\phi_{\rm ini} \ll 1$, the asymptotic conditions correspond to 
\begin{equation}
    k(\phi\to0) \to k_0,
    \qquad
    q(\phi\to0) \to0.
    \label{eq:ACG_asymptotics}
\end{equation}

In this work, however, we drop this assumption and consider the wider Rolling Galileon theory space spanned by $k(\phi)$ and $q(\phi)$, so that the coupling coefficients may evolve over all of cosmic history. We proceed by presenting the theory in a convenient frame, in which these invariants appear most naturally.

\subsection{The Canonical Braiding Frame}

A particularly convenient frame choice is the \textit{canonical braiding frame}, defined by requiring the transformed braiding prefactor to be unity, 
\begin{equation}
    \widetilde B = B \varphi_\phi^3 = 1
    \qquad\implies \quad
    \varphi_\phi = B^{-1/3}.
    \label{eq:ACG_cubic_norm}
\end{equation}
In this frame, $\widetilde B$ is a constant and hence $\widetilde B_\phi=0$. It then follows from~\eqref{eq:first_invariant} and~\eqref{eq:second_invariant} that
\begin{equation}
    \widetilde k(\phi) = \widetilde A(\phi),
    \qquad
    \widetilde q(\phi) = \widetilde C(\phi),
\end{equation}
and the invariant functions coincide directly with the coefficients of the action. 
Substituting into~\eqref{eq:general_transformed_ACG}, we obtain 
\begin{equation}
    \mathcal L_\phi
    =
    - k(\phi) X
    - X\Box \phi
    + q(\phi) X^2,
    \label{eq:ACG_action}
\end{equation}
where, for notational simplicity, we have now dropped the tildes. 


Other scalar frames are equally valid, most notably the more standard canonical kinetic frame. We adopt the canonical braiding frame in the main text as a convention tailored around the braiding strength---the physical sector under investigation. By using the braiding strength as the normalising scale, we implicitly assume that braiding always exists, whilst the rolling functions $k$ and $q$ are the kinetic and quadratic coefficients measured in braiding units. Indeed, the canonical kinetic frame gives an equivalent invariant description, which we derive in App.~\ref{app:peril_of_canonical_kinetic_frame}. It is also more convenient for constructing simple models in terms of a directly evolving braiding coefficient, as we demonstrate in App.~\ref{app:minimal_ansatz}.



\section{Phenomenological Conditions}
\label{sec:phantom_crossing}
Having characterised the closed two-function theory space through the invariant pair $\{k(\phi), q(\phi)\}$, we now ask what rolling behaviour is selected by three requirements: i) a late-time phantom crossing, ii) a positive ISW signature, and iii) a healthy screened scalar force in cosmic voids. We derive analytical conditions that translate into phenomenologically motivated constraints on $k$ and $q$, thereby restricting the functional freedom within this theory space. We begin by recasting the scalar sector, together with a possible cosmological constant, as an effective dark energy fluid.

\subsection{Effective Fluid Description}
Rolling Galileons~\eqref{eq:ACG_action} sit within the minimally coupled luminal Horndeski class ($G_4=M_\mathrm{Pl}^2/2$)~\cite{Kobayashi:2019hrl}, 
\begin{equation}
    \mathcal L_\phi = K(\phi,X)  -G_3(\phi,X)\Box\phi,
    \label{eq:Horndeski_functions}
\end{equation}
with 
\begin{equation}
    K(\phi,X) = -k(\phi)X + q(\phi)X^2,
    \quad
    G_3(\phi,X) = X.
    \label{eq:K_G3_IDK}
\end{equation}
On a spatially flat Friedmann-Lemaitre-Robertson-Walker (FLRW) background,
\begin{equation}
    ds^2 = -dt^2 + a^2(t) d\mathbf x^2 ,
    \label{eq:FLRW_metric}
\end{equation}
where $a(t)$ is the scale factor, $H\coloneqq \dot a/a $ is the Hubble rate, and overdot denotes differentiation with respect to time $t$. The scalar kinetic term consequently reduces to $X=\dot \phi^2/2$.

The scalar sector then admits an effective fluid description, with energy density and pressure\footnote{Note these correspond to $\tilde{\mathcal E}$ (A.1) and $\tilde{\mathcal P}$ (A.2) in~\cite{Bellini_2014}.}
\begin{align}
    \rho_\phi &= -k(\phi)X + 3q(\phi)X^2 + 6H\dot\phi  X, 
    \label{eq:rho_phi}
    \\
    p_\phi   &= -k(\phi)X + q(\phi)X^2 - 2X\ddot\phi,
    \label{eq:p_phi}
\end{align}
Treating the cosmological constant as a perfect fluid with $\rho_\Lambda = -p_\Lambda = \Lambda$, the total dark energy sector becomes $\rho_{\rm DE} = \rho_\phi + \Lambda$ and $p_{\rm DE} = p_\phi - \Lambda$, with an equation of state
\begin{equation}
    w_{\rm DE}
    = \frac{p_\phi - \Lambda}{\rho_\phi + \Lambda}.
    \label{eq:wDE_def}
\end{equation}
Introducing the fractional scalar contribution to dark energy, 
\begin{equation}
    f_\phi(z)
    \coloneqq \frac{\Omega_\phi(z)}{\Omega_\phi(z) + \Omega_\Lambda(z)}
    = \frac{\rho_\phi(z)}{\rho_{\rm DE}(z)},
    \label{eq:fphi}
\end{equation}
allows~\eqref{eq:wDE_def} to be rewritten as 
\begin{equation}
    w_{\rm DE} = f_\phi (1 + w_\phi) - 1,
    \label{eq:wDE_fphi}
\end{equation}
where $w_\phi\coloneqq p_\phi/\rho_\phi$. The departure of the total dark energy sector from the phantom divide is thus the scalar departure $1+w_\phi$, weighted by the fractional contribution $f_\phi$. 
The location of the crossing is set by the scalar sector, with $f_\phi$ controlling the departing magnitude. 
Combining~\eqref{eq:rho_phi} and~\eqref{eq:p_phi}, the scalar sector is governed by
\begin{equation}
    \rho_\phi + p_\phi
    = 2X \left( -k + 2qX + 3H\dot\phi - \ddot\phi\right),
    \label{eq:rho_plus_p}
\end{equation}
whose vanishing marks the crossing. 

Note that although~\eqref{eq:rho_plus_p} may become negative, the total source of curvature $T_{\mu\nu} = T_{\mu\nu}^\mathrm{(m)} + T_{\mu\nu}^\mathrm{(DE)}$ can remain NEC-respecting. 
The perturbative stability of the scalar mode is ensured by requiring a positive scalar sound speed, $c_s^2>0$, and no ghosts, $\mathcal D>0$, which for Rolling Galileons are given by~\cite{Bellini_2014}
\begin{align}
    c_s^2
    &=
    \frac{2\left(-\dot H + H X\dot\phi  -2X^3- 3X\ddot\phi \right)
    +
    \rho_m+p_m}{2X\mathcal{D}},
    \label{eq:cs2}
    \\
    \mathcal{D} 
    &=
    6H\dot\phi-k+6X(q+X).
    \label{eq:ghostly}
\end{align}
As shown in Fig.~\ref{fig:stability}, these are shown to be valid for the Rolling Galileon models confronted with the data.



\subsection{Phantom Crossing} 
\label{subsec:phantom_crossing}
It is convenient to define
\begin{equation}
    \mathcal{P}
    \coloneqq -k + q \dot\phi^2 + 3H\dot\phi - \ddot\phi,
    \label{eq:off-shell_crossing_condition}
\end{equation}
so that $\rho_\phi + p_\phi = 2X \mathcal P$ and, by~\eqref{eq:wDE_fphi},
\begin{equation}
    w_{\rm DE} + 1
    = \frac{\dot \phi^2 f_\phi}{\rho_\phi}\mathcal P .
\end{equation}
The total dark energy sector thus falls into three regimes:
\begin{equation}
\mathcal P  
\begin{cases}
      < 0 &\implies w_\phi < -1 \quad (\textit{phantom}),\\
      = 0 &\implies w_\phi = -1 \quad (\textit{marginal}),\\
      > 0 &\implies w_\phi > -1 \quad (\textit{non-phantom}),
\end{cases}
\nonumber
\end{equation}
with a genuine phantom-to-non-phantom crossing requiring $\dot{\mathcal P} > 0$ at the crossing point.

To isolate the dependence of $\mathcal P$ on the invariant functions $k(\phi)$ and $q(\phi)$, we substitute the scalar equation of motion~\cite{Bellini_2014}
\begin{equation}
    \ddot\phi
    =
    \frac{
    3H(k - q \dot\phi^2)\dot\phi
    - 3\left(3H^2 + \dot H - \frac16 k_\phi\right)\dot\phi^2
    - \frac34 q_\phi \dot\phi^4
    }{
    6H\dot\phi - k + 3q \dot\phi^2
    },
    \label{eq:phiddot_ACG}
\end{equation}
into~\eqref{eq:off-shell_crossing_condition}, yielding
\begin{align}
    \mathcal P
    =& -k + q\dot\phi^2 + 6H\dot\phi 
    \nonumber \\
    & - \frac{12(3+\epsilon_H)H^2\dot\phi^2 + 24Hq\dot\phi^3 + \left(2k_\phi - 3q_\phi\dot\phi^2\right)\dot\phi^2}{4\left(6H\dot\phi -k + 3q\dot\phi^2\right)},
    \label{eq:P_exact}
\end{align}
where $\epsilon_H \coloneqq -\dot H/H^2>0$. The first three terms are the algebraic quantities from the kinetic and braiding sectors, whilst the fraction collects the remaining background and derivative corrections, with the second bracket isolating the explicit field-dependence contributions.

Eq.~\eqref{eq:P_exact} shows that $\{k,q\}$ control phantom crossing in three places: the algebraic combination $(-k+q\dot\phi^2)$, the derivative bracket $(-2k_\phi+3q_\phi\dot\phi^2)$, and the denominator $(-k+3q\dot\phi^2)$. The dominant requirement is to reduce the algebraic contribution by allowing $k$ to decrease as the field rolls. This also contributes favourably through the derivative term, and raises the denominator since it enters there with the same sign. A positive $q$ further assists through the same algebraic combination and also increases the denominator,
helping to prevent the denominator from becoming singular.
Its effect is nevertheless subordinate because it is always weighted by powers of $\dot\phi^2$ and is therefore small whilst the field is thawing. Thus, phantom crossing primarily selects $k_\phi<0$, and mildly favours $q>0$ with some freedom to shape its specific behaviour by the further phenomenological constraints to which we now turn.

\subsection{The ISW Effect}
\label{subsec:isw}

The Integrated Sachs-Wolfe (ISW) effect~\cite{1967_Sachs_Wolfe} occurs when CMB photons traverse gravitational potentials that evolve with time. It is therefore sensitive not only to the amplitude of the lensing potential, but also to how quickly that amplitude changes. If the lensing potential decays, as occurs in the $\Lambda$CDM model, the sign of the cross-correlation between the ISW and galaxies is positive; it is negative if the potential deepens. Analyses to date strongly support a positive ISW-galaxy cross-correlation signal~\cite{Renk:2017rzu, Seraille:2024beb}. The ISW sign can therefore be a powerful tool for ruling out modified theories of gravity.

On linear scales, scalar perturbations may be written in the Newtonian gauge as
\begin{equation}
    ds^2 = -(1+2\Phi)dt^2 + a^2(t) (1-2\Psi) d\mathbf x^2 ,
    \label{eq:newtonian_gauge_metric}
\end{equation}
with $\Phi$ and $\Psi$ the scalar gravitational potentials. The Newtonian potential is sourced through the modified Poisson equation
\begin{equation}
    \nabla^2 \Phi
    = \frac32 H_0^2 \frac{\Omega_{m0}}{a}  \mu(a) \delta,
    \label{eq:Sigma_def}
\end{equation}
where $\mu$ measures the linear modification to the gravitational strength felt by matter. The Weyl-potential modification may be similarly written as
\begin{equation}
    \nabla^2(\Phi+\Psi)
    = 3 H_0^2 \frac{ \Omega_{m0}}{a}  \Sigma(a) \delta.
    \label{eq:Sigma_def}
\end{equation}
The GR limit corresponds to $\mu=\Sigma=1$.\linkedfootnotemark{assumptions}

For minimally coupled luminal Horndeski theories, the growth and lensing modifications coincide, $\mu=\Sigma$. For Rolling Galileons, this gives
\begin{equation}
    \mu=\Sigma
    =  1+\frac{\dot\phi^4}{H^2 \mathcal D c_s^2},
    \label{eq:mu_Sigma_RG}
\end{equation}
so the no-ghost $\mathcal D>0$ and gradient-stability $c_s^2>0$ conditions guarantee $\Sigma>1$. The same quantity $\mu$ controls the linear growth factor $D_1$, which, with $D_1(z=0)=1$, obeys
\begin{equation}
    D_1'' + \left(2+ \frac{E'}{E} \right) D_1'
    - \frac{3}{2} \Omega_m(a) \mu(a)  D_1
    = 0,
    \label{eq:growth_D1}
\end{equation}
where primes denote derivatives with respect to $\ln a$, and $E\coloneqq H/H_0$. 
The linear growth rate and time variation of the lensing modification is then respectively measured by 
\begin{equation}
    f_1 \coloneqq \frac{d \ln D_1}{d\ln a},
    \qquad
    \zeta\coloneqq \frac{d\ln }{d\ln a} \left( \frac{\Sigma(a)}{\Sigma(a=1)} \right).
    \label{eq:zeta_def}
\end{equation}
\linkedfootnotetext{assumptions}{We work to linear order in perturbations and under the subhorizon quasistatic limit \cite{Boisseau:2000pr,Esposito-Farese:2000pbo,Copeland:2006wr,Tsujikawa:2007gd} and the weak-field approximation \cite{Fidler:2017pnb}. Additionally, we assume that the scalar sector introduces no effective mass scale parametrically larger than $H$, allowing the modifications considered here to be treated as functions of time only~\cite{Baker:2014zva}. For a derivation of these equations beyond those assumptions, we refer to \cite{Sirera:2026klo}.}
These affect the ISW strength via the integral\footnote{See Eq.~(46) of~\cite{HiCOLA:Krishna.paper}.}
\begin{equation}
    S_{\rm ISW}
    = \int d\eta 
    D_1^2(a) \Sigma (a) E(a)  \left[ 1 - f_1(a) - \zeta(a) \right], 
    \label{eq:S_ISW}
\end{equation}
where $\eta$ is the comoving distance. Since $D_1^2\Sigma E$ is positive, we impose the integrated condition
\begin{equation}
    \int d\eta\,\mathcal I(a)>0,
    \qquad 
    \mathcal I(a) \coloneqq 1 - f_1(a) - \zeta(a).
\end{equation}
We have verified numerically that this condition ensures $S_{\rm ISW}>0$ for all cosmologies considered.

\subsection{Modified Force in Voids} 
\label{subsec:modifed_force}

Scalar-tensor theories generically mediate an additional fifth force upon matter. 
Local tests of gravity, however, strongly constrain such forces~\cite{Will:2014kxa}, and so must be suppressed, or \textit{screened}, in dense environments. In Galileon-type theories this is achieved via the Vainshtein mechanism, whereby the derivative self-interactions dominate near massive sources, suppressing the fifth force within the Vainshtein radius $r_*$ and recovering GR locally~\cite{Vainshtein:1972sx, Babichev:2013usa,  Kimura:2011dc}. This mechanism is what allows the cubic Galileon to be viable in dense, non-relativistic environments. 

This same mechanism, however, carries a known liability within the opposite regime. In underdense environments, the fifth force risks becoming ill-defined---the real solution can cease to exist~\cite{Baker:2018mnu}. This problem has long been encountered by N-body simulations of Vainshtein-screened theories, where it is typically handled by an ad hoc prescription for the force inside voids whenever the solution turns imaginary~\cite{Barreira:2013eea, Winther:2015pta, Moretti:2026dfz,Moretti:2026axy}. 
Following~\cite{Moretti:2026axy}, which promotes the reality of the screened force in underdense configurations to a theory-level consistency condition, we impose this requirement on the Rolling Galileon theory space and derive the corresponding model-specific constraint.

The screened scalar force implemented in \texttt{Hi-COLA}~\cite{Wright:2022krq, Gupta:2024seu, HiCOLA:Krishna.paper} takes the form
\begin{equation}
    \frac{F_\phi}{F_N}
    = 2\mathcal C  \frac1\chi \left[\sqrt{1 + \chi} - 1\right],
    \label{eq:Fphi}
\end{equation}
where, in Vainshtein-type theories, $\chi \coloneqq (r_*/r)^3$ is the cubed Vainshtein-radius ratio. The Vainshtein radius $r_*$ is set by the enclosed mass perturbation $\delta M$ via $r_*^3 \propto \delta M / H^2$. 
Assuming a uniform spherical overdensity, $\delta M = \delta\rho_m \frac{4\pi}{3}r^3$, and writing the fractional density contrast $\delta_m \coloneqq \delta\rho_m/\bar{\rho}_m$ with $\bar{\rho}_m =3H_0^2\Omega_m / (8\pi G_N a^3)$ from the Friedmann equation, $\chi$ becomes proportional to the matter density contrast~\cite{Winther:2014cia}. 
For Rolling Galileons with~\eqref{eq:K_G3_IDK}, this proportionality takes the form 
\begin{equation}
    \chi = \left(\frac{\mathcal U(a)  H\dot\phi}{2 \mathcal W}\right)^{2} \delta_m,
    \label{eq:chi_delta}
\end{equation}
where we have defined
\begin{align}
    \mathcal U(a)^2 &\coloneqq E(a) \Omega_m(a),
    \\
    \mathcal W &\coloneqq \ddot\phi + 2H\dot\phi - \frac12\left(k - q\dot\phi^2\right) - \frac14\dot\phi^4.
    \label{eq:W_def}
\end{align}
We find numerically that $\mathcal W>0$ along all cosmologies considered; note that $\chi/\delta_m$ diverges where this combination vanishes, so its sign is necessarily fixed. In numerical implementations, more realistic density profiles may be approximated via Gaussian smoothing, but this does not automatically remove the pathology~\cite{Wright:2022krq}.


In overdensities $\delta_m > 0$, so one has $\chi > 0$ and~\eqref{eq:Fphi} is regular. In voids, however, $\delta_m < 0$ renders $\chi$ negative, and the root becomes imaginary once $|\chi|$ exceeds unity. 
Absence of pathologies in voids therefore requires
\begin{equation}
    \frac{\chi}{\delta_m} \leq 1,
    \label{eq:chi_delta_condition}
\end{equation}
throughout the cosmologically relevant interval. 

We note that this condition rests on the assumption of pure Vainshtein screening. Rolling Galileons, with field-dependent kinetic operator coefficients and a quadratic term, generically have screening of mixed type~\cite{Sirera:2026klo}. The interplay of the multiple mechanisms may indeed modify the behaviour of the force in voids, perhaps even helping to alleviate this pathology. A full characterisation of this with the master equation~\cite{Sirera:2026klo}, however, lies beyond the scope of the present work.

Since $\mathcal U$, $H$, $\dot \phi$, and $\mathcal W$ are all positive in the cosmologies considered, the viability condition~\eqref{eq:chi_delta_condition} applied to~\eqref{eq:chi_delta} is equivalent to $\mathcal F \geq 0$, where
\begin{align}
    \mathcal F
    &\coloneqq \mathcal W - \frac12 \mathcal U(a) H\dot\phi
    \nonumber \\
    &= -\frac12 \left(k - q\dot\phi^2 \right) + H \dot\phi\left(2 - \frac12 \mathcal U(a)\right) - \frac14 \dot\phi^4 + \ddot\phi .
    \label{eq:Q_implicit}
\end{align}
To identify the dependence on the invariant pair $\{k, q\}$ (as per Eqs.~\eqref{eq:first_invariant} and~\eqref{eq:second_invariant}), we proceed exactly as for the phantom crossing in Sec.~\ref{subsec:phantom_crossing}. Substituting the scalar field equation~\eqref{eq:phiddot_ACG} for $\ddot \phi$ yields
\begin{align}
    \mathcal F
    =&  -\frac12\left(k - q\dot\phi^2\right) - H \dot\phi\left(1 + \frac12\mathcal U(a)\right) - \frac14 \dot\phi^4
    \nonumber\\
    &+ \frac{12 (3+\epsilon_H) H^2 \dot\phi^2 + 24 H q \dot\phi^3 + \left(2k_\phi - 3q_\phi \dot\phi^2 \right)\dot\phi^2}{4\left(6H\dot\phi -k + 3q\dot\phi^2\right)}.
    \label{eq:Q_exact}
\end{align}
We see that $\{k,q\}$ enter in the same three places as the phantom crossing quantity $\mathcal P$ \eqref{eq:P_exact}. A decreasing $k$ and positive $q$ help decrease the first algebraic term $(k - q\dot\phi^2)$. Similarly, this helps keep the denominator of the fraction regular. However, as $\ddot \phi$ enters $\mathcal F$ with the opposite sign to $\mathcal P$, $k_\phi<0$ actually suppresses $\mathcal F$ and endangers the voids. Thus the rolling profile of a single function is caught in tension between phantom crossing and void viability.

But of course, we still have yet to select the rolling profile of $q$. Whilst an increasing $q$ would further aid the phantom crossing condition, it would equally worsen the void condition. Thus we may use it as a natural counterbalance: taking $q_\phi<0$ shifts the trade-off toward the void condition. Void viability therefore selects a positive but decaying quadratic-to-braiding strength. Collectively, then, we have the phenomenological rolling profile:
\begin{equation}
    \dot\phi > 0 : \quad k_\phi < 0, \quad q > 0, \quad q_\phi < 0.
    \label{eq:preferred}
\end{equation}

This also has the benefit of an intuitive interpretation. Since $k$ and $q$ measure the relative kinetic and quadratic-kinetic strengths with respect to the braiding strength, their simultaneous decrease simply implies an increasing braiding strength. 
The simplicity of this rolling mechanism---an increasing braiding strength---is exploited in App.~\ref{app:minimal_ansatz}, where we derive the minimal Rolling Galileon model analysed numerically in the following section.

\section{Numerical Exploration of a Minimal Ansatz} 
\label{sec:carpets}
The conditions~\eqref{eq:preferred} 
restrict the qualitative behaviour of the pair $\{k,q\}$ (Eqs.~\eqref{eq:first_invariant} and~\eqref{eq:second_invariant}). To test whether these conditions can actually be met simultaneously, we introduce a minimal functional ansatz for $\{k,q\}$ and map the resulting viable region of parameter space. We then confront the model with current data to locate the data-preferred region relative to that viable sector.

\subsection{\texttt{Hi-COLA}: The Numerical Solver}
\label{sec:Hi-COLA}
To explore the Rolling Galileon theory space numerically, we use \texttt{Hi-COLA}~\cite{Wright:2022krq, Gupta:2024seu, HiCOLA:Krishna.paper}, a package for fast, approximate simulations of structure formation in luminal Horndeski gravity. It comprises two building blocks: first, a symbolic Python front-end that takes a user-specified Lagrangian, constructs the corresponding field equations, and solves for the background expansion, linear growth, and screened fifth force; and second, a modified $N$-body \texttt{COLA} solver back-end~\cite{Tassev:2013pn}. In the present work, we use only the front end of \texttt{Hi-COLA}. 

A run is specified by the Lagrangian and a small set of cosmological parameters: the reference Hubble constant $H_0^\mathrm{ref}$, which corresponds to the $\Lambda$CDM back-scaled expansion used to set the high-redshift initial conditions; the physical baryon and cold dark matter densities $\omega_b$ and $\omega_c$; and the initial fraction $f_\phi^\mathrm{ini}$ of the total dark energy density carried by the scalar field, which fixes the residual cosmological-constant contribution. The radiation sector is held fixed at standard values and is not varied. Since the modified expansion generically yields $E(z)|_{z=0}=H(z)/H_0|_{z=0}\neq1$, the actual $H_0$ derived in a Horndeski model is a derived quantity, related to $H_0^\mathrm{ref}$ by a rescaling and differing from it in general. 
For further details, we refer to the companion paper~\cite{HiCOLA:Krishna.paper}.

For Lagrangians with explicit $\phi$-dependence, the user fixes either the initial field value $\phi_\mathrm{ini}$ or its derivative $\phi'_\mathrm{ini}=d\phi/d\ln a|_{a_\mathrm{ini}}$, with the remaining quantity determined self-consistently by the Friedmann (closure) constraint at the initial time. With Rolling Galileons, this closure condition is quartic in $\dot\phi$ and admits up to four roots. We typically find two complex roots and two real roots of opposite sign, corresponding to physically distinct branches. We retain the positive root, on which the minimal ansatz realises the intended phenomenology.

\begin{figure*}[t] 
    \centering
    \includegraphics[width=1.0\linewidth]{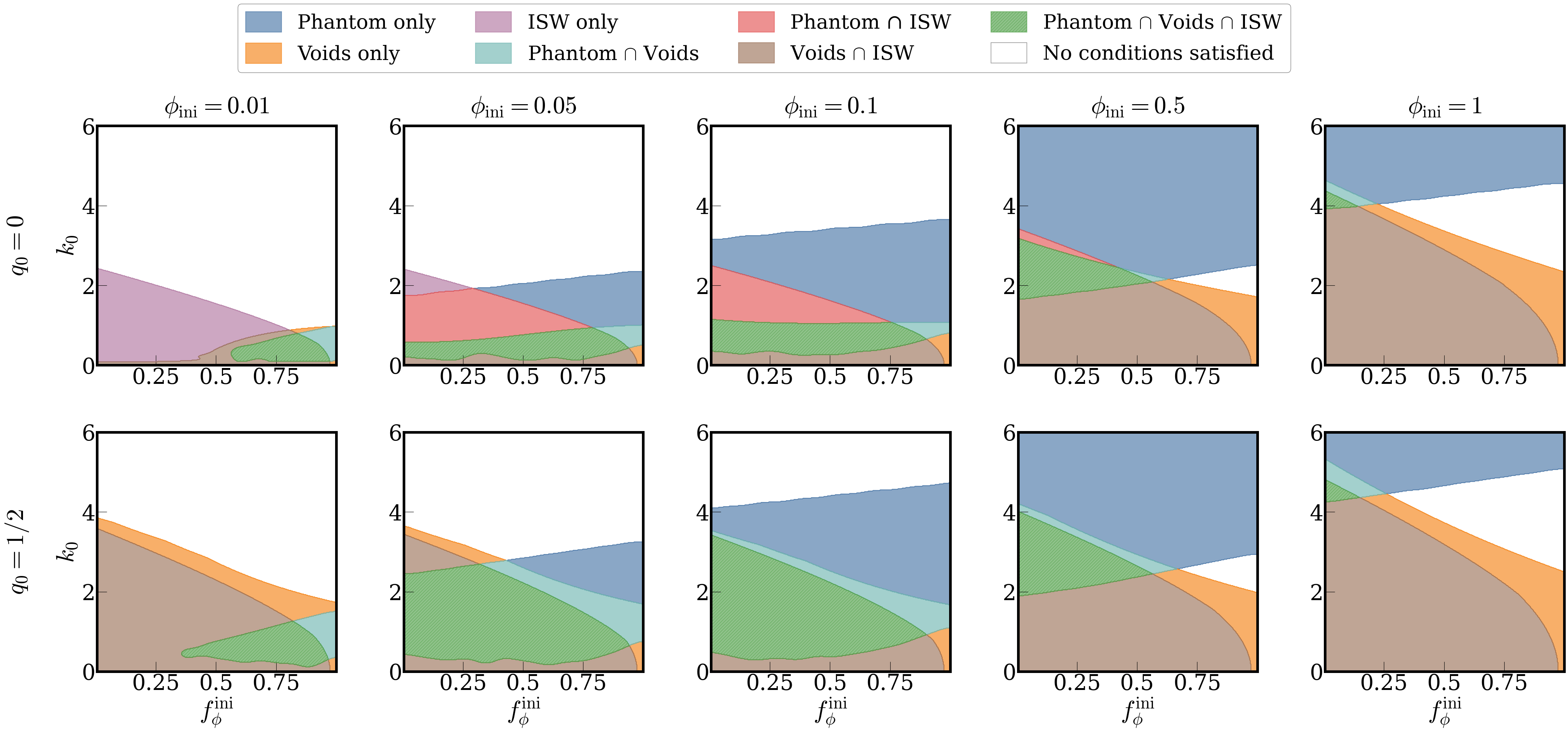}
    \caption{Carpet plots of cosmological viability across $\{f_\phi,  k_0\}$ parameter space for the $q_0=0$ slice (top row) and the $q_0=1/2$ slice (bottom row) of the simple rolling theory~\eqref{eq:minimal_theory}, evaluated at five slices of the initial field value $\phi_{\rm ini}$ as indicated. At each grid point, the model is classified according to which subset of the three viability criteria it satisfies: phantom crossing, a positive ISW signature, and healthy modified force in voids. Coloured regions indicate where the individual criteria and their intersections are satisfied, as shown in the legend. The green region marks the fully viable sector. 
    We see that viable regions of Rolling Galileon models exist, with the $q_0=1/2$ slice offering larger regions of viable parameter space.}
    \label{fig:carpets}
\end{figure*}


\subsection{The Minimal Functional Ansatz}

To evaluate a specific model, we require explicit expressions for the invariant pair $\{k(\phi), q(\phi)\}$. 
In App.~\ref{app:minimal_ansatz}, we construct the simplest one-function rolling extension of the cubic Galileon in canonical kinetic frame by allowing the braiding coefficient to depend linearly on the field. This links the otherwise distinct invariants through $q(\phi)\propto k(\phi)^2$, and in the canonical braiding frame~\eqref{eq:ACG_action}, the simple model 
is described by
\begin{equation}
    k(\phi)=\frac{k_0}{\sqrt{\phi}},
    \qquad
    q(\phi)=\frac{q_0}{\phi}.
    \label{eq:minimal_theory}
\end{equation}
For $k_0, q_0>0$ and $\dot \phi>0$, both functions are positive and decrease as the field rolls, corresponding to an increasing braiding strength relative to the kinetic sector---i.e. the phenomenologically preferred rolling profile~\eqref{eq:preferred}.

The minimal rolling theory~\eqref{eq:minimal_theory} introduces four free parameters $\{f_\phi^\mathrm{ini}, k_0, q_0, \phi_{\rm ini}\}$, 
two more than the shift-symmetric cubic Galileon before imposing the tracker condition. Here, $f_\phi^\mathrm{ini}$ sets the initial scalar contribution to dark energy; the constants $k_0$ and $q_0$ control the amplitudes of the initial kinetic and quadratic strengths relative to the braiding strength, and $\phi_\mathrm{ini}$ fixes their initial value.

\subsection{Cosmological Viability}
\label{sec:viable_space}

To isolate and demonstrate the roles of the invariant pair $\{k(\phi),q(\phi)\}$, we first analyse the minimal theory~\eqref{eq:minimal_theory} in two configurations. To reduce the dimensionality of the parameter space from $4$ to $3$, we consider two illustrative slices, fixing the quadratic amplitude to $q_0=0$ and $q_0=1/2$. The former isolates the effects of a positive and decreasing kinetic-to-braiding strength, and the latter demonstrates the effect of having an additional positive and decreasing quadratic-to-braiding term.



For each model, we sweep the parameter space spanned by $f_\phi^\mathrm{ini} \in [0,1]$ and $k_0 \in [0, 6]$ on a $51\times51$ grid at fixed initial field $\phi_{\rm ini}\in\{0.01, 0.05, 0.1, 0.5, 1\}$. These five values are somewhat arbitrary, chosen from a wider exploratory scan to find, at roughly logarithmic intervals, a range over which the phantom crossing, ISW, and void-viable regions have significant overlap. At each grid point the full background is integrated using \texttt{Hi-COLA}, and the resulting cosmology is classified according to whether it
\begin{enumerate}
    \item crosses the phantom divide at $z<1$ (see Sec.~\ref{subsec:phantom_crossing});
    \item maintains a positive ISW signature (see Sec.~\ref{subsec:isw});
    \item has a healthy modified force in voids (see Sec.~\ref{subsec:modifed_force}). 
\end{enumerate}
We identify the cosmologically viable region as the simultaneous intersection of all three criteria. We present the resulting classification as a \textit{carpet plot},\footnote{So named by Ashim Sen Gupta for their visual resemblance to carpet designs of the 1960s.} identifying the regions selected by each individual diagnostic and their mutual overlap across several slices of $\phi_{\rm ini}$. 
 
The results are shown in Fig.~\ref{fig:carpets}, with the $q_0=0$ slice of~\eqref{eq:minimal_theory} in the top row and the $q_0=1/2$ slice in the bottom row.
Both models display a qualitatively similar phantom crossing band. For intermediate values of $\phi_\mathrm{ini}$, this band extends across the entire $f_\phi^\mathrm{ini}$ range but is confined to an interval of $k_0$. This interval shifts to larger $k_0$ as $\phi_\mathrm{ini}$ increases and smaller $k_0$ as it decreases, eventually disappearing from the sampled range for sufficiently small $\phi_\mathrm{ini}$. This trend is almost unaffected by the presence of $q$, in agreement with the analytic understanding of Sec.~\ref{sec:phantom_crossing}, where $k$ drives the crossing with $q$ having little effect.

The ISW-viable region is likewise similar in both models, occupying the lower-left of the carpet whose viability requires $f_\phi<1$ throughout, so a nonzero cosmological constant contribution is always necessary.

The void condition is where the two models differ significantly. In the $q_0=0$ slice, the void-viable region is confined to a narrow band, expanding only as $\phi_{\rm ini}$ is made sufficiently large. The three-way intersection, i.e. the viable region, is consequently small. In the $q_0=1/2$ slice, by contrast, the void-viable region expands to overlap almost entirely with the ISW region, leaving only a narrow ridge between them. When the phantom crossing strip overlaps, it therefore falls almost automatically within a jointly viable region. 

Having identified numerically the distinct phenomenological roles of the kinetic-to-braiding $k$ and quadratic-to-braiding $q$ strengths, we now confront this simple model with current data, asking where the observationally preferred parameter region lies with respect to the viability space and, in particular, whether it falls within the fully viable sector. 

\subsection{Constraints from Observational Data}
\label{subsec:data}

The carpet plots of the preceding subsection mapped the viable sector at fixed $H_0^{\rm ref}$ and $\phi_{\rm ini}$, revealing that $k$ controls the phantom crossing whilst $q$ mainly enlarges the void-viable region. We now free these parameters and confront the model with data, asking where the preferred region falls relative to the viability conditions. We fit the expansion history inferred from the compressed Planck~\cite{Planck2020}, DESI DR2~\cite{DESI:2025zgx}, and DES-Dovekie~\cite{Popovic2025} likelihoods, sampling the posterior with \texttt{emcee}~\cite{Foreman2013}, which is well suited to the non-Gaussian posteriors encountered here. The likelihoods and sampling pipeline are those of the companion paper~\cite{HiCOLA:Krishna.paper}, into which RG is implemented. Here, we summarise only the model-specific elements. Following~\cite{HiCOLA:Krishna.paper}, we apply an a posteriori positive-ISW selection 
enforcing a non-negative ISW strength \eqref{eq:S_ISW} in place of a full galaxy-ISW likelihood. By contrast, we do not impose any prior associated with the screened force in voids. We sample two configurations of the minimal theory. The first is its $k$-only restriction
\begin{equation}
    k(\phi) = \frac{k_0}{\sqrt\phi},
    \qquad
    q(\phi) = 0,
    \label{eq:k-only_model}
\end{equation}
and the second is the full $k$-$q$ model~\eqref{eq:minimal_theory}. We therefore sample $\{H_0^\mathrm{ref}, \omega_b, \omega_c, f_\phi^\mathrm{ini},\phi_\mathrm{ini},k_0, q_0\}$
. 
\newpage

The resulting constraints are shown in Fig.~\ref{fig:corner}, App.~\ref{app:model_indep_constraints}, and summarised in Table~\ref{tab:bestfit}. In both models, $\omega_b$ and $\omega_c$ are essentially unshifted and the derived $H_0 \simeq 67.3$-$67.8$ $\mathrm{km\,s^{-1}\,Mpc^{-1}}$, 
recovering the standard $\Lambda$CDM background parameters. 
The modification therefore resides entirely in the dark energy sector. For the $k$-only model, the posterior is well localised. For the $k$-$q$ model, the posterior is concentrated toward a large quadratic-to-braiding strength ($q_0$), although the amplitude is strongly degenerate with $k_0$ and $\phi_{\rm ini}$. However,
the physical histories are much more tightly constrained than the bare parameters. Indeed, both RG models improve the best-fit expansion history relative to $\Lambda$CDM, with $\Delta\chi^2_\mathrm{MAP}\simeq-11$. The improvement is already captured by the $k$-only model. Restoring $q$ changes the best-fit value only slightly, $\Delta\chi^2_\mathrm{MAP}\simeq-0.5$, preserving the same good fit whilst moving the posterior into the void-healthy region.



As for the fitted histories, we see in the top-left panel of Fig.~\ref{fig:predictive} that both models drive the desired phantom-to-quintessence dark energy crossing, occurring around $z\sim0.75$. We also find that the ISW stays positive across the posterior by construction, both sitting below the $\Lambda$CDM line (top-right panel). Moreover, both models predict a relatively small modification to the effective gravitational strength (lower-left panel), with $\mu(z=0)\sim 1.12$-$1.17$, and slightly lower for the $k$-$q$ model. The resulting changes to the growth rate are therefore expected to remain small.
\newpage

\begin{plainwidetext}
\begin{center}

    \refstepcounter{figure}
    \label{fig:corner}

    \begin{minipage}[t]{0.49\textwidth}
        \centering
        \includegraphics[width=\linewidth]
            {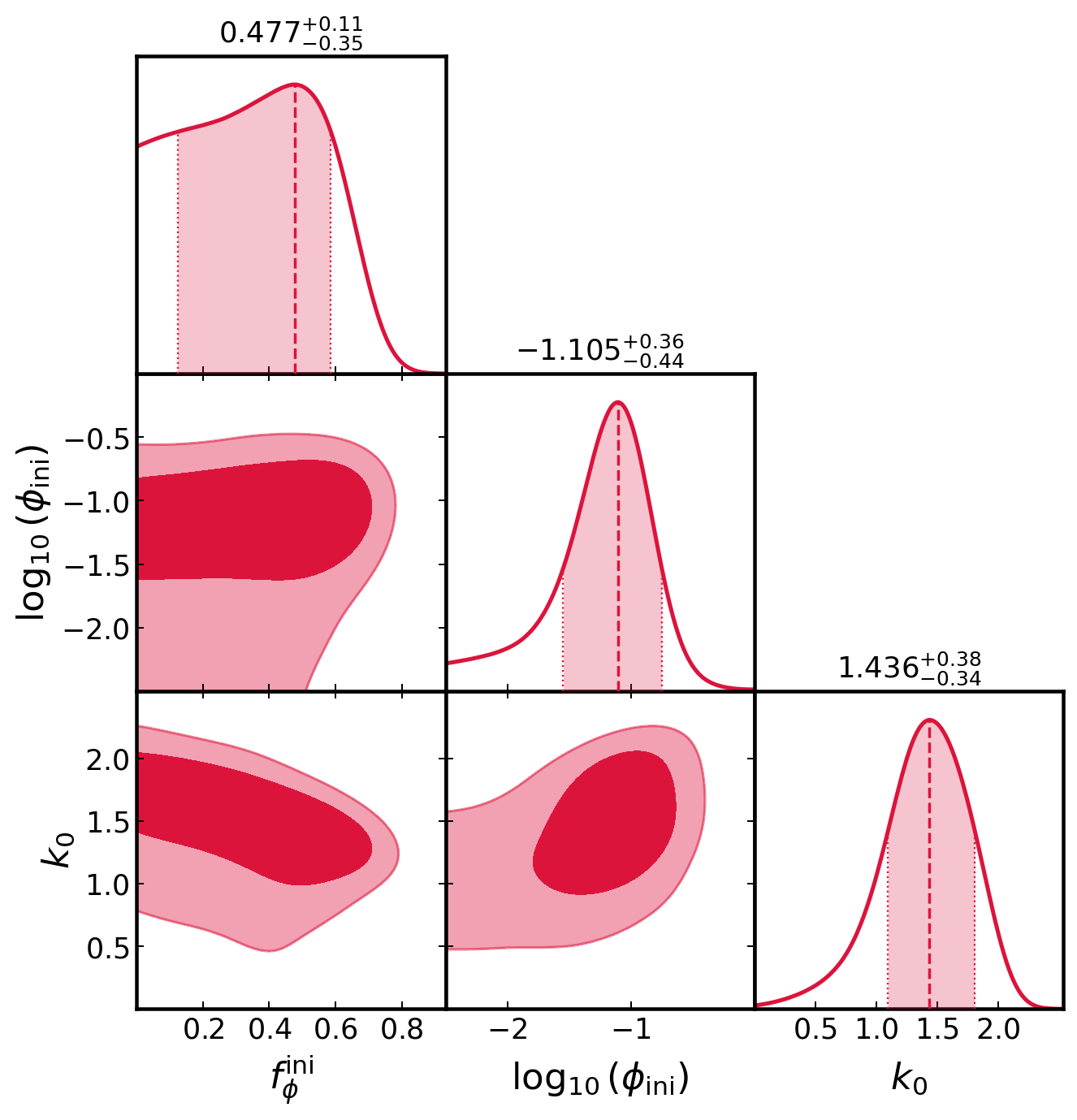}

        \refstepcounter{subfigure}
        \label{fig:corner-konly}
        \small\rmfamily
        (\alph{subfigure}) $k$-only model.
    \end{minipage}
    \hfill
    \begin{minipage}[t]{0.49\textwidth}
        \centering
        \includegraphics[width=\linewidth]
            {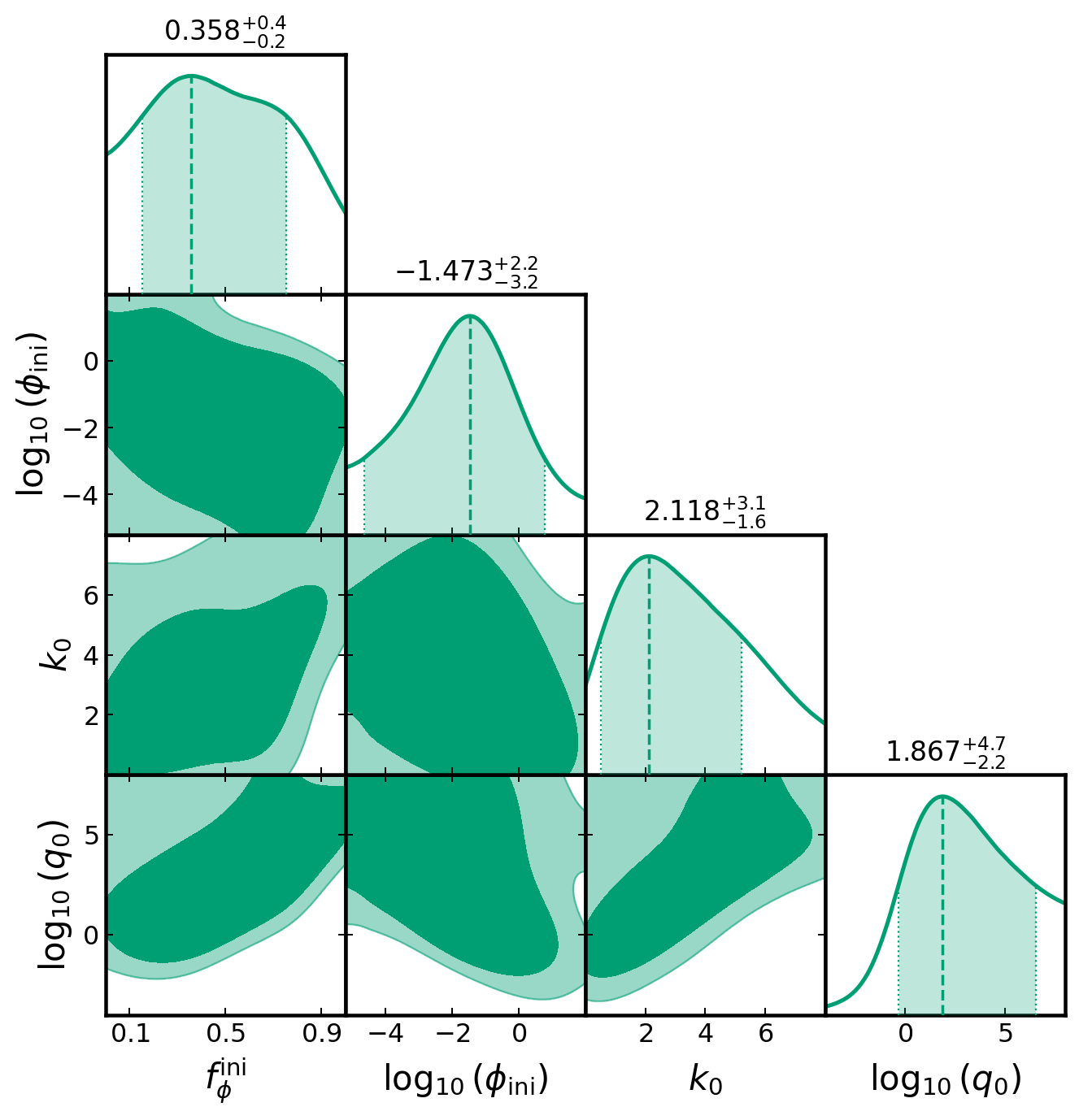}

        \refstepcounter{subfigure}
        \label{fig:corner-kq}
        \small\rmfamily
        (\alph{subfigure}) $k$-$q$ model.
    \end{minipage}

    \vspace{0.5em}

    \begin{minipage}{\textwidth}
        \small\rmfamily
        \raggedright
        \parindent=0pt
        FIG.~\thefigure:
        Marginalised posterior constraints for the two minimal Rolling Galileon models fitted to the joint CMB+BAO+SN data set with the positive-ISW condition imposed. The left panel shows the $k$-only model~\eqref{eq:k-only_model}, whilst the right panel shows the full minimal $k$-$q$ model~\eqref{eq:minimal_theory}. All parameter priors are listed in Table~\ref{tab:bestfit}, and the model-independent parameters are presented in App.~\ref{app:model_indep_constraints}. 
    \end{minipage}

\end{center}
\end{plainwidetext}


\begin{plainwidetext}
\begin{center}
    \includegraphics[width=1\textwidth]{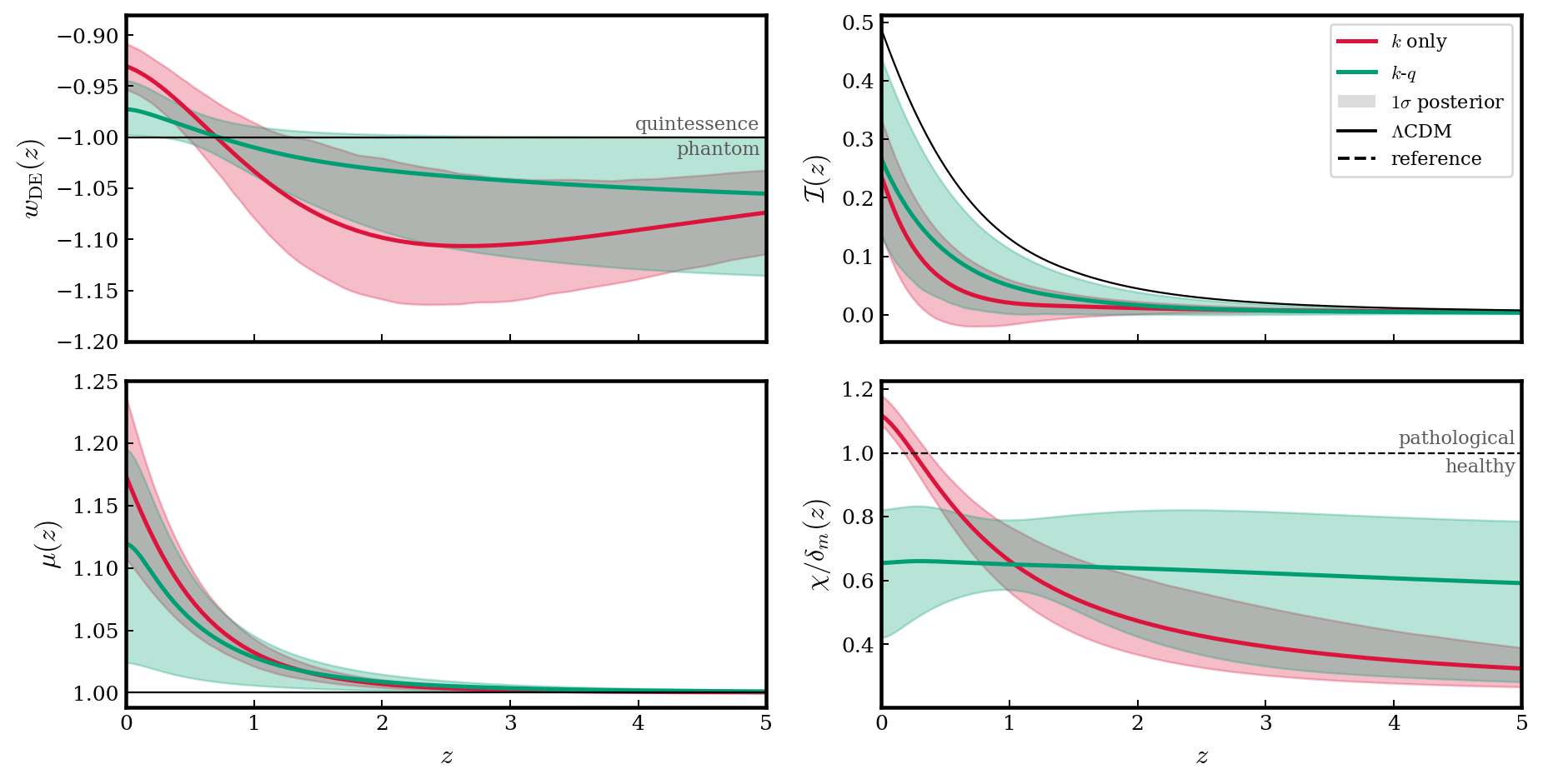}

    \refstepcounter{figure}
    \label{fig:predictive}
    \begin{minipage}{1\textwidth}
    \small\rmfamily
    \flushing
    \parindent=0pt
    FIG.~\thefigure: 
    Posterior predictive viability diagnostics (see Sec.~\ref{sec:phantom_crossing}), together with the effective gravitational strength $\mu$~\eqref{eq:Sigma_def}, for the $k$-only (red) and the $k$-$q$ (green) Rolling Galileon models. Solid lines denote the mean histories and shaded regions the $1\sigma$ posterior band. The dashed lines mark a specific reference point for the respective quantity, as so labelled. 
    Both models achieve the desired phantom crossing and satisfy the positive ISW selection, with a fairly weak modification to the effective gravitational strength: the mean histories give $\mu(z=0)\simeq1.12$-$1.17$.
    The full $k$-$q$ additionally remains below the void-pathology threshold, whereas the $k$-only model violates this condition at low redshift.  
    \end{minipage}
\end{center}
\end{plainwidetext} 

\textcolor{white}{Hello there.} \clearpage

\begin{table}[h!]
\centering
\setlength{\tabcolsep}{4pt} 
\footnotesize  

\begin{tabular}{@{}lcccc@{}}
\toprule\toprule
& \multicolumn{2}{c}{$k$-only}
& \multicolumn{2}{c}{$k$-$q$} \\
\cmidrule(lr){2-3}\cmidrule(lr){4-5}
Parameter & Prior & Constraint & Prior & Constraint \\
\midrule\midrule
$H_0^\mathrm{ref}$
    & $[30,200]$ & $66.23^{+12.5}_{-5.86}$
    & $[30,200]$ & $67.37^{+12.4}_{-12.8}$ \\[0.2em]
$100\omega_b$
    & $[2,2.5]$ & $2.236\pm0.013$
    & $[2,2.5]$ & $2.236\pm0.014$ \\[0.2em]
$100\omega_c$
    & $[10,15]$ & $11.804\pm0.097$
    & $[10,15]$ & $11.796\pm0.090$ \\[0.2em]
$f_\phi^\mathrm{ini}$
    & $[0,1]$ & $0.4772^{+0.108}_{-0.354}$
    & $[0,1]$ & $0.3578^{+0.396}_{-0.204}$ \\[0.2em]
$\log_{10}\phi_\mathrm{ini}$
    & $[-3,1]$ & $-1.105^{+0.357}_{-0.444}$
    & $[-10,3]$ & $-1.867^{+2.25}_{-3.17}$ \\[0.2em]
$k_0$
    & $[0,6]$ & $1.436^{+0.375}_{-0.341}$
    & $[0,100]$ & $2.118^{+3.09}_{-1.60}$ \\[0.2em]
$\log_{10}q_0$
    & $-$ & $-$
    & $[-10,10]$ & $1.867^{+4.66}_{-2.21}$ \\[0.2em]
\midrule
$H_0$
    & $-$ & $67.33\pm0.57$
    & $-$ & $67.88^{+0.504}_{-0.549}$ \\[0.2em]
$f_\phi^0$
    & $-$ & $0.2935^{+0.1044}_{-0.1064}$
    & $-$ & $0.5135^{+0.3664}_{-0.3949}$ \\[0.2em]
$\log_{10}\phi'_\mathrm{ini}$
    & $-$ & $-6.07^{+0.11}_{-0.16}$
    & $-$ & $-6.06^{+0.12}_{-0.17}$ \\[0.2em]
$z_\mathrm{cross}$
    & $-$ & $0.705^{+0.595}_{-0.228}$
    & $-$ & $0.725^{+1.524}_{-0.479}$ \\[0.2em]
$\max\mu$
    & $-$ & $1.1734^{+0.0648}_{-0.0653}$
    & $-$ & $1.1197^{+0.0760}_{-0.0959}$ \\[0.2em]
$\min\mathcal I$
    & $-$ & $0.0045^{+0.0009}_{-0.0249}$
    & $-$ & $0.0035^{+0.0031}_{-0.0044}$ \\[0.2em]
$\max(\chi/\delta_m)$
    & $-$ & $1.1177^{+0.0618}_{-0.0325}$
    & $-$ & $0.6604^{+0.1717}_{-0.0903}$ \\[0.2em]
$\min c_s^2$
    & $-$ & $0.2743^{+0.0133}_{-0.0218}$
    & $-$ & $0.1930^{+0.1004}_{-0.1182}$ \\[0.2em]
$\min\mathcal D$
    & $-$ & $0.0064^{+0.0039}_{-0.0042}$
    & $-$ & $0.0853^{+0.1035}_{-0.0820}$ \\[0.2em]
\midrule
$\Delta\chi^2_{\rm MAP}$
    & \multicolumn{2}{c}{$-11.2$}
    & \multicolumn{2}{c}{$-11.7$} \\
\bottomrule\bottomrule
\end{tabular}
\caption{Marginalised parameter constraints for the $k$-only and full $k$-$q$ models, from the joint CMB+BAO+SN fit with the positive-ISW condition.
The upper block lists sampled parameters; the middle block the derived parameters. The lower block gives the goodness of the fit relative to $\Lambda$CDM, evaluated at the respective maximum a posteriori (MAP) values, $\Delta\chi^2_\mathrm{MAP} = \chi^2_\mathrm{RG,MAP} - \chi^2_{\Lambda\mathrm{CDM,MAP}}$.}
\label{tab:bestfit}
\end{table}

\begin{figure*}[t!]
    \centering
    \includegraphics[width=1\linewidth]{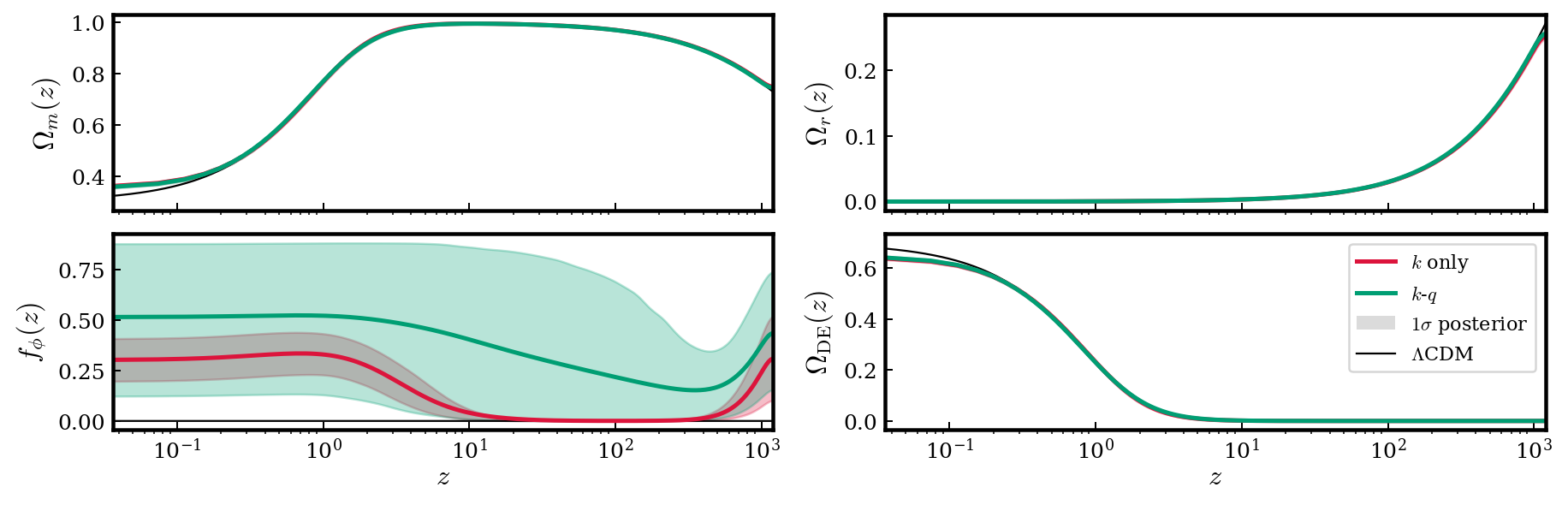}
    \caption{Posterior background evolution for the $k$-only (red) and $k$-$q$ (green) Rolling Galileon models.
    The $\Lambda$CDM reference (black) is nearly indistinguishable from the Rolling Galileon density fractions, with only a small separation visible at the lowest redshift as the rolling scalar modifies the late-time expansion.
    The density parameters $\Omega_m(z)$, $\Omega_r(z)$, and $\Omega_{\rm DE}(z)$ are therefore tightly constrained, with the residual model freedom confined primarily to the internal composition of the dark energy sector, quantified by the fractional scalar contribution $f_\phi(z)$.
    }
    \label{fig:background}
\end{figure*}

The remaining background quantities are presented in Fig.~\ref{fig:background}. The density parameters $\Omega_m(z)$ and $\Omega_r(z)$, and hence the total dark energy abundance $\Omega_\mathrm{DE}(z)$, are tightly constrained (see Table~\ref{tab:bestfit} and Fig.~\ref{fig:corner_other}) and track their $\Lambda$CDM histories, consistent with the standard background parameters recovered in Table~\ref{tab:bestfit}. The residual freedom lies not in how much dark energy there is, but in what it is made of. This internal composition is measured by $f_\phi(z)$. The data prefer a genuinely hybrid dark energy sector, with both a rolling scalar and a surviving constant vacuum component, but are relatively indifferent to the precise split between them. This is reflected in Fig.~\ref{fig:background}, where the total expansion history is tightly fixed whilst the scalar fraction retains significant freedom, and in the non-Gaussian posterior of Fig.~\ref{fig:corner}(b), where $f_\phi^\mathrm{ini}$ remains only weakly localised.

\begin{figure}[h!]
    \centering
    \includegraphics[width=1\linewidth]{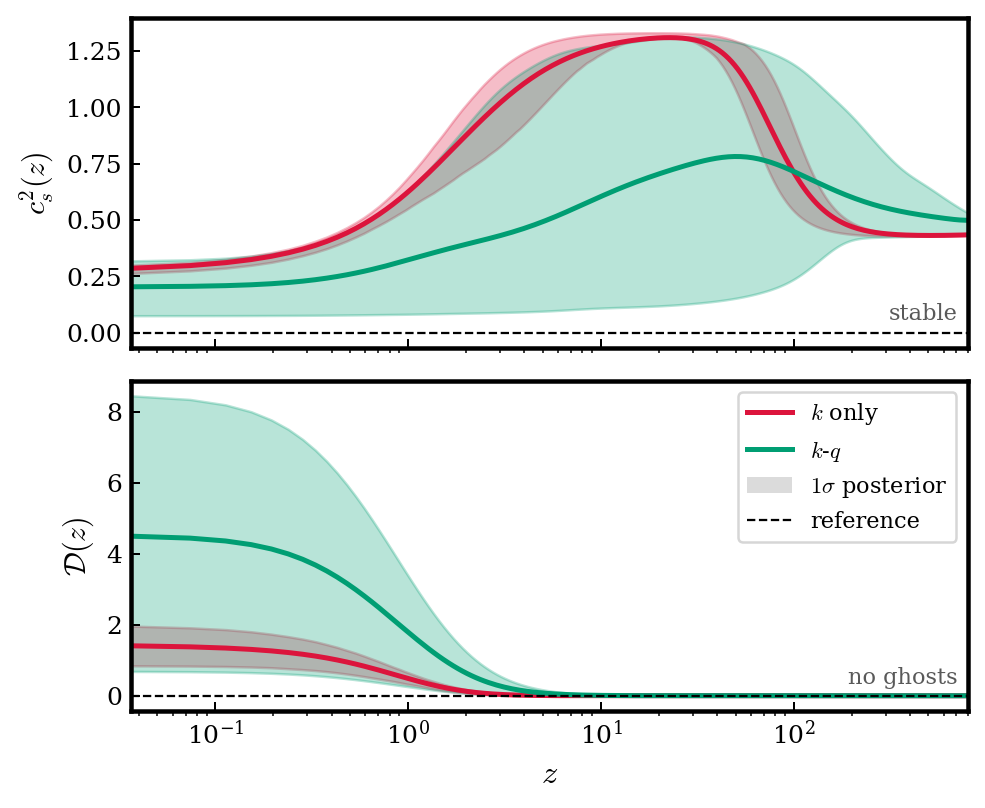}
    \caption{Perturbative stability check of the scalar field on an FLRW background in the fitted histories. Both models remain free of ghosts and gradient instabilities, satisfying $c_s^2>0$ \eqref{eq:cs2} and $\mathcal D>0$ \eqref{eq:ghostly} throughout.
    }
    \label{fig:stability}
\end{figure}

The void condition is where the two models truly diverge, and where the analytic argument of the void counterbalance in Sec.~\ref{sec:phantom_crossing} is realised. In the $k$-only model, the screened-force ratio is driven above the pathology threshold $\chi/\delta_m = 1$ at low redshift---see the lower-right panel of Fig.~\ref{fig:predictive}---implying the breakdown of the Vainshtein solution in voids that likewise afflicts the ACG models of the companion paper~\cite{HiCOLA:Krishna.paper}. Restoring the quadratic sector is found to alleviate this issue. In the full $k$-$q$ model, $\chi/\delta_m$ remains well below unity across all redshifts, with the mean and $1\sigma$ band remaining comfortably healthy. Hence, a positive, decaying $q$ relieves the void tension left by $k$ alone. As there was no imposed prior, this result is a genuine posterior prediction selected by the data.

We also verify the perturbative stability of the scalar field on the FLRW background in Fig.~\ref{fig:stability}. Throughout the fitted histories, $c_s^2>0$~\eqref{eq:cs2} excludes gradient instabilities and $\mathcal D>0$~\eqref{eq:ghostly} ensures the absence of ghosts. Interestingly, the $k$-only model is driven through a period of superluminal propagation, whereas the full $k$-$q$ model may remain subluminal throughout. 





\begin{figure*}[t!]
    \centering
    \includegraphics[width=1\linewidth]{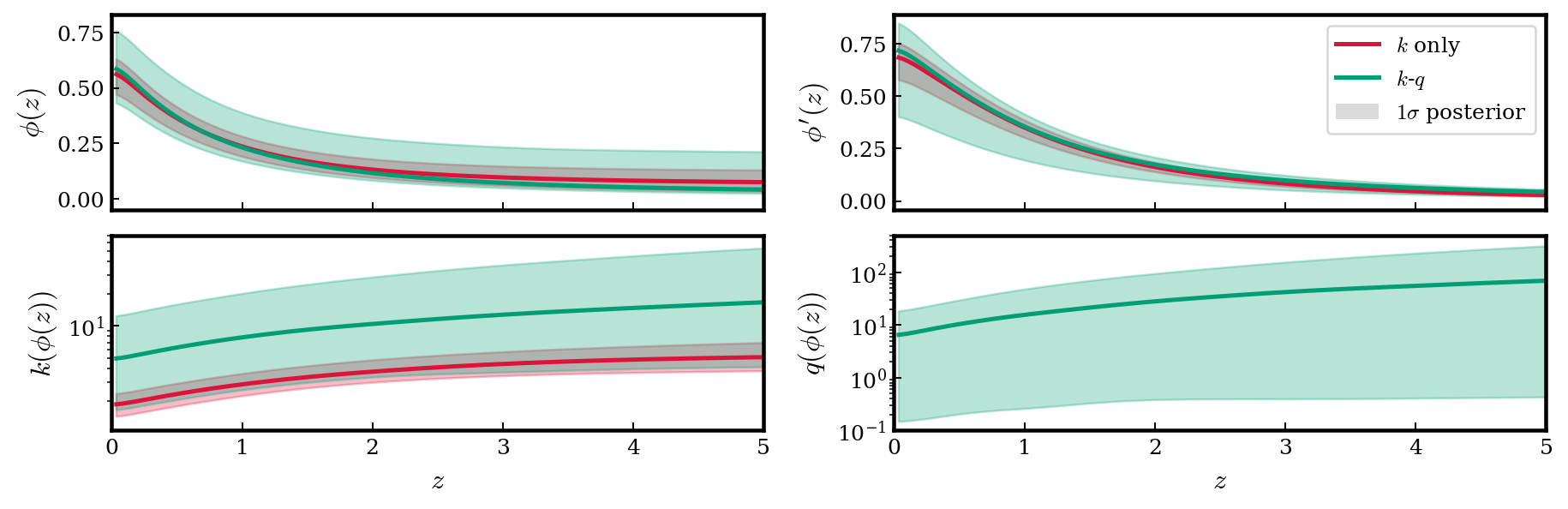}
    \caption{Posterior evolution of the scalar field and its first derivative, together with the functions $\{k,q\}$, given by~\eqref{eq:first_invariant} and~\eqref{eq:second_invariant}, for the $k$-only model (red), defined by~\eqref{eq:k-only_model}, and $k$-$q$ model (green), defined by~\eqref{eq:minimal_theory}. As the dark energy density grows, the field thaws and rolls to larger $\phi$, driving $k$ and $q$ downwards. 
    }
    \label{fig:phi}
\end{figure*}

The corresponding field histories are shown in Fig.~\ref{fig:phi}. At early times the scalar is held near-frozen. As the dark energy density grows, the field thaws, activating the rolling phase of the Galileon. The relative weighting between the kinetic, quadratic, and braiding terms redistributes as intended, with $k(\phi)$ and $q(\phi)$ both decreasing at recent redshift $z<3$. It is this late-time evolution of the operator weights that drives the phenomenological behaviour of Fig.~\ref{fig:predictive}. It is also observed that, provided $\phi_\mathrm{ini}$ is sufficiently small, the field thaws onto effectively the same trajectory, so that the late-time history is insensitive to its precise value. This accounts for the pronounced tails towards small values of the marginalised $\phi_\mathrm{ini}$ (Fig.~\ref{fig:corner}).

The overall Rolling Galileon behaviour preferred by the data is therefore: matter dilutes, the dark energy sector grows, the field thaws and begins to roll, and the relative braiding strength increases, driving the effective dark energy contribution across the phantom divide.




\section{Conclusions}
\label{sec:conclusions}

In this work, we introduced Rolling Galileons (RG) as the minimal shift-symmetry breaking generalisation of the cubic Galileon in which the coupling coefficients roll with the field. The full theory space, closed under field redefinitions, is fully characterised by two field redefinition-invariant functions, $k(\phi)$ and $q(\phi)$. We derived analytical conditions under which an increasing braiding strength yields a stable phantom crossing, a positive ISW signature, and a healthy screened force in cosmic voids. 
We further confirmed this preference through MCMC sampling, with the $4$ distinct Rolling Galileons studied in this work and in~\cite{HiCOLA:Krishna.paper} collected in Table~\ref{tab:RG_results}.
They therefore constitute simple and observationally favoured dynamical dark energy models.

The alternative approaches are presented in Table~\ref{tab:model_comparison}, and each carries a cost. Non-minimal coupling fits the expansion well~\cite{Ye:2024ywg, Wolf:2024stt, Wolf:2024eph, Gu:2025xie}, though the joint fit of distances and growth warrants caution~\cite{Garcia-Garcia:2026nzy, Linder:2025zxb}. Within minimally coupled braiding, the divide cannot be crossed whilst the shift symmetry remains intact, even when admitting a non-standard kinetic structure~\cite{Linder:2025pqt}; the symmetry must therefore be broken, but breaking it with a scalar potential also brings problems with growth~\cite{Garcia-Garcia:2026nzy, Wolf_2026}. 

Rolling Galileons require no such field-dependent potential. We nevertheless retain a residual cosmological constant contribution throughout ($f_\phi<1$), in light of the difficulties encountered by no-$\Lambda$ potential models~\cite{Calderon:2026hbr}. A field-dependent potential could of course be added, but its contribution enters the phantom and void viability conditions with opposite signs. It therefore cannot, by itself, alleviate both sides of the phantom-void tension, as shown in App.~\ref{app:potential}.

The same pathology appears in the Asymptotically Cubic Galileon (ACG) models of~\cite{HiCOLA:Krishna.paper}, whose Rolling Galileon descriptions we derive in App.~\ref{app:ckf}. Each breaks the symmetry through a single rolling function, crosses the phantom divide, and fits the background data, but their vanishing or negative quadratic-to-braiding strength, $q(\phi)\le0$, leads to the breakdown of the Vainshtein solution in voids. 
Thus, whilst all four models in Table~\ref{tab:RG_results} improve upon $\Lambda$CDM by $\Delta\chi^2_\mathrm{MAP}\simeq-11$, only the $k$-$q$ model~\eqref{eq:minimal_theory} simultaneously satisfies all three conditions. 

Interestingly, the viable $k$-$q$ histories also satisfy the qualitative behaviour recently argued to ease the apparent preference for negative effective neutrino masses: a phantom phase at higher redshift, followed by a later crossing~\cite{Yang:2026yaq}. This suggests that a joint analysis with $\sum m_\nu$ free would be a good follow-up test of the model.

\begin{table}
\centering
\renewcommand{\arraystretch}{1.}
\setlength{\tabcolsep}{6pt}

\footnotesize
\begin{tabular}{p{2.9cm} c c c c}
\hline\hline
Model & PC & ISW & Voids & $\Delta\chi^2_\mathrm{MAP}$ \\[0.4em]
\hline
Growing $\mathcal{G}(\phi)$~\cite{HiCOLA:Krishna.paper}
  & \cmark & \cmark & \xmark & $-10.5$ \\ 
  Decaying $\mathcal{K}(\phi)$~\cite{HiCOLA:Krishna.paper}
  & \cmark & \cmark & \xmark & $-10.6$ \\ 
  $k$-only model \eqref{eq:k-only_model}
  & \cmark & \cmark & \xmark & $-11.2$ \\ 
  $k$-$q$ model \eqref{eq:minimal_theory}
  & \cmark & \cmark & \cmark & $-11.7$ \\
\hline\hline
\multicolumn{5}{@{}p{\columnwidth}@{}}{%
  \footnotesize $^{\ddagger}$ Imposed as a diagnostic condition in the MCMC;
  see Sec.~\ref{subsec:data}.%
} \\
\end{tabular}
\caption{The diagnostics of Sec.~\ref{sec:phantom_crossing}, applied to the ACG models of the companion paper~\cite{HiCOLA:Krishna.paper} and the RG models of this work, where "PC" stands for phantom crossing. The $\Delta\chi^2_\mathrm{MAP}$ is quoted relative to $\Lambda$CDM for the joint CMB$+$BAO$+$SN$+\![\text{ISW}]^{\ddagger}$ fit. 
}
\label{tab:RG_results}
\end{table}






But of course, several caveats should be noted. Foremost, the void condition~\eqref{eq:chi_delta_condition} rests on the derivation for a uniform spherical approximation, so whether the low-redshift breakdown of the $k$-only model reflects a genuine failure of the Vainshtein solution or merely the limit of these approximations must ultimately be settled by a full nonlinear treatment in underdense environments. The screening of Rolling Galileons may, moreover, be of mixed type rather than purely Vainshtein, and its full characterisation via the master equation for screening~\cite{Sirera:2026klo} remains to be done. Our quantitative conclusions, furthermore, pertain to the simple $k$-$q$ model~\eqref{eq:minimal_theory} rather than to the Rolling Galileon space as a whole. We merely demonstrated its potential through one of the simplest models it permits. 
Finally, the ISW condition imposed in the MCMC should be viewed as a proof-of-concept prior on the ISW sign, rather than a replacement for a full ISW likelihood. Ultimately, one should combine background distances, linear growth and lensing observables, and nonlinear structure formation in the same pipeline. These questions, however, are left for future work.

\section{Acknowledgements} 
\label{sec:acknowledgements}
It gives us much pleasure in giving thanks to Ashim Sen Gupta, Kazuya Koyama, and Eric Linder 
for helpful comments and discussions. 
J.~Hallam is supported by a PhD studentship from UKRI-STFC. 
T.~Baker, K.~Naidoo and S.~Sirera are supported by ERC Starting Grant SHADE (grant no. StG 949572). T.~Baker is further supported by a Royal Society University Research Fellowship (grant no. URF\textbackslash R\textbackslash 231006). 

\appendix
\section{The ACG Models as Rolling Galileons}
\label{app:ckf}
The companion paper~\cite{HiCOLA:Krishna.paper} introduces two Asymptotic Cubic Galileon (ACG) models--Growing $\mathcal G(\phi)$ and Decaying $\mathcal K(\phi)$--defined in a frame distinct from the canonical braiding frame of the main text. To locate these models within the Rolling Galileon theory space, we first give the general rules for passing between frames, then apply them to each ACG model in turn, closing with a comparison.

\subsection{General Frame Transformations}
The invariant pair $\{ k,q\} $ may be extracted from any scalar-field frame. Given the general action 
\begin{equation}
    \mathcal L_\varphi = 
    -A(\varphi)\mathcal X 
    - B(\varphi) \mathcal X\Box\varphi 
    + C(\varphi)\mathcal X^2,
    \label{eq:general_RG_action}
\end{equation}
the invariants, expressed initially as functions of the original field coordinate $\varphi$, are
\begin{equation}
    k(\varphi) = \frac{A(\varphi)}{B(\varphi)^{2/3}}, \qquad
    q(\varphi) = \frac{C(\varphi) - \frac{2}{3}B_\varphi(\varphi)}{B(\varphi)^{4/3}}.
    \label{eq:app_invariants}
\end{equation}
They may then be expressed in the canonical braiding frame, defined by a unit braiding coefficient $1 = \widetilde B(\phi) = B(\varphi(\phi))\varphi_\phi^3$, by introducing the field $\phi$ through
\begin{equation}
    \frac{d\varphi}{d\phi}
    =B(\varphi(\phi))^{-1/3}
    \quad\iff\quad
    \frac{d\phi}{d\varphi}=B(\varphi)^{1/3}.
    \label{eq:app_cbf_map}
\end{equation}
After integrating this relation and substituting $\varphi=\varphi(\phi)$, the functions $k(\phi)=k(\varphi(\phi))$ and $q(\phi)=q(\varphi(\phi))$ appear directly as the action coefficients in~\eqref{eq:ACG_action}. We fix the irrelevant integration constant in $\phi$ by taking $\phi=0$ at $\varphi=0$.

\subsection{Growing $\mathcal G(\phi)$}
The first model introduced in \cite{HiCOLA:Krishna.paper}, the Growing $\mathcal G$ model, reads
\begin{equation}
    \mathcal L_\varphi
    = -k_1 \mathcal X - g_{31}(1+c_{g3}\varphi)\mathcal X\Box\varphi.
\end{equation}
Comparing with the general action~\eqref{eq:general_RG_action} gives
\begin{equation}
 A(\varphi)=k_1,
 \quad
 B(\varphi)=g_{31}(1+c_{g3}\varphi),
 \quad
 C(\varphi)=0,
\end{equation}
and hence $B_\varphi=g_{31}c_{g3}$. Eq.~\eqref{eq:app_invariants} therefore gives 
\begin{align}
    k(\varphi)
    &= \frac{k_1}{\left[g_{31}(1+c_{g3}\varphi)\right]^{2/3}},
    \label{eq:growing_G_k_varphi}
    \\
    q(\varphi)
    &= -\frac{2c_{g3}}
    {3 g_{31}^{1/3}(1+c_{g3}\varphi)^{4/3}
    }
    .
    \label{eq:growing_G_k-q_varphi}
\end{align}
To pass to the canonical braiding frame, we use the inverse form of~\eqref{eq:app_cbf_map},
\begin{equation}
    \frac{d\phi}{d\varphi}
    =
    g_{31}^{1/3}(1+c_{g3}\varphi)^{1/3}.
\end{equation}
Integrating and imposing $\phi=0$ at $\varphi=0$ gives 
\begin{equation}
    \phi
    =
    \frac{3g_{31}^{1/3}}{4c_{g3}}
    \left[
    (1+c_{g3}\varphi)^{4/3}-1
    \right].
\end{equation}
Defining $\phi_0\coloneqq 4c_{g3}/3g_{31}^{1/3}$, this may be inverted as 
\begin{equation}
    1 + c_{g3} \varphi
    = 
    (1+ \phi_0 \phi)^{3/4} 
\end{equation}
Substituting this combination into \eqref{eq:growing_G_k_varphi} and \eqref{eq:growing_G_k-q_varphi}, utilising the invariance of the functions $\tilde f(\phi) = f(\varphi(\phi))$, and defining $k_0\coloneqq k_1 g_{31}^{-2/3}$, yields
\begin{equation}
    \widetilde k(\phi) = \frac{k_0}{\sqrt{1+\phi_0\phi}},
    \qquad
    \widetilde q(\phi) = -\frac{\phi_0}{2(1+\phi_0\phi)}.
    \label{eq:app_growing_G_kq}
\end{equation}

\subsection{Decaying $\mathcal K(\phi)$}
The second model introduced in~\cite{HiCOLA:Krishna.paper}, the decaying $\mathcal K$ model, is given by
\begin{equation}
    \mathcal L_\varphi
    = -k_1 e^{-c_k\varphi}\mathcal X - g_{31} \mathcal X\Box\varphi.
\end{equation}
Thus
\begin{equation}
    A(\varphi) = k_1 e^{-c_k\varphi},
    \quad
    B(\varphi) = g_{31},
    \quad
    C(\varphi)=0.
\end{equation}
Since $B_\varphi=0$, no quadratic coefficient is induced, and 
\begin{equation}
    k(\varphi) = \frac{k_1}{g_{31}^{2/3}}e^{-c_k\varphi},
    \qquad q(\varphi) = 0.
\end{equation}

Because $B$ is constant, the field transformation is simply
\begin{equation}
    \frac{d \phi}{d \varphi} = g_{31}^{1/3}
    \quad\implies\quad
    \phi = g_{31}^{1/3} \varphi,
\end{equation}
where the field origins again coincide. Substitution therefore gives
\begin{equation}
    \widetilde k(\phi) = k_0 e^{-\phi_0\phi},
    \qquad 
    \widetilde q(\phi) = 0,
    \label{eq:app_decaying_K_kq}
\end{equation}
where $k_0 \coloneqq k_1 g_{31}^{-2/3}$ and $\phi_0 \coloneqq c_k g_{31}^{1/3}$.

\subsection{Comparison}
Equations~\eqref{eq:app_growing_G_kq} and~\eqref{eq:app_decaying_K_kq} tell us both the similarities and differences between the two ACG models studied in~\cite{HiCOLA:Krishna.paper}. In both cases, with $0<\phi_\mathrm{ini}\ll1$, $k(\phi)$ decreases monotonically from a positive constant when the field starts rolling. With a square-root rational decay for Growing $\mathcal G$ and exponential decay for Decaying $\mathcal K$, they admit qualitatively similar profiles for $k(\phi)$. Since $k(\phi)$ is the main driver of the phantom crossing, both models are expected to yield similar histories. 

The models are nonetheless genuinely distinct. Growing $\mathcal G$ carries $q(\phi)<0$, whilst Decaying $\mathcal K$ has $q(\phi)=0$ identically. Thus despite the fact that both models have the desired rolling kinetic-to-braiding strength, they still lie outside the void-viable region of the Rolling Galileon theory space which preferred $q>0$ and $q_\phi<0$, as identified in Sec.~\ref{sec:phantom_crossing}. Indeed, the companion paper~\cite{HiCOLA:Krishna.paper} confirms that both these ACG models exhibit void pathologies ($\chi/\delta_m>1$) at low redshift.

\section{The Canonical Kinetic Frame}
\label{app:peril_of_canonical_kinetic_frame}

For completeness, we present an equivalent parametrisation of the two-function theory space adapted to the canonical kinetic frame. The main text instead uses the canonical braiding frame because the Rolling Galileon construction is organised around the evolving braiding sector: the functions $k$ and $q$ then measure the kinetic and quadratic strength directly in braiding units. The two frames differ only by a choice of field coordinate.


We may define the canonical kinetic field $\psi$ through
\begin{equation}
    \varphi_\psi = A(\varphi(\psi))^{-1/2},
    \quad\iff\quad
    \psi_\varphi = A(\varphi)^{1/2}.
    \label{eq:kinetic_proper_field}
\end{equation}
The transformed kinetic coefficient is therefore $\widetilde A(\psi) = A(\varphi(\psi))\varphi_\psi^2=1$. As with the braiding coordinate, we must specify the patch on which we reside. If $A$ vanishes or changes sign, $\psi$ ceases to provide a valid coordinate there. We thus restrict ourselves to the $A>0$ patch, which reduces to the convention $a>0$ in the constant-coefficient theory~\eqref{eq:general_CG}. 

Using $\varphi_{\psi\psi} = -A_\varphi/(2A^2)$, the remaining transformed coefficients are
\begin{align}
    g(\psi) 
    &\coloneqq 
    \widetilde B(\psi) 
    = \frac{B(\varphi)}{A(\varphi)^{3/2}}\Bigg|_{\varphi=\varphi(\psi)}
    \label{eq:kinetic_g}
    \\
    h(\psi)
    &\coloneqq
    \widetilde C(\psi)
    =
    \frac{A(\varphi)C(\varphi) -B(\varphi) A_\varphi(\varphi)}
    {A(\varphi)^3}\Bigg|_{\varphi=\varphi(\psi)}.
    \label{eq:kinetic_h}
\end{align}
Thus $g$ and $h$ are respectively the \textit{braiding-to-kinetic} and \textit{quadratic-to-kinetic} strengths.

By construction, the kinetic-adapted functions $\{g,h\}$ appear directly as the remaining coefficients of the action, since $\widetilde A_\psi=g_\psi=0$, 
\begin{equation}
    \mathcal L_\psi
    = -\mathcal X  -g(\psi) \mathcal X  \Box\psi + h(\psi)\mathcal X^2,
    \label{eq:app_kinframe}
\end{equation}
where $\mathcal X\equiv - \frac12 g^{\mu\nu} \partial_\mu \psi \partial_\nu \psi$ here.

To relate this pair to the braiding-adapted functions $\{k(\phi),q(\phi)\}$~\eqref{eq:first_invariant} and ~\eqref{eq:second_invariant}, where $\phi$ is the canonical braiding field, we consider the $B>0$ patch used in the main text. The canonical braiding coordinate satisfies $g(\psi) \psi_\phi^3=1$, or equivalently $\psi_\phi = g(\psi)^{-1/3}$. Substituting into the expressions for the braiding-adapted functions~\eqref{eq:first_invariant} and ~\eqref{eq:second_invariant} gives
\begin{align}
    k(\phi)&=\frac1{g(\psi(\phi))^{2/3}},
    \label{eq:k_from_g}
    \\[0.2em]
    \qquad
    q(\phi)&=
    \frac{h(\psi(\phi))-\frac{2}{3}g_\psi(\psi(\phi)) }{g(\psi(\phi))^{4/3}}.
    \label{eq:kinetic_braiding_inverse}
\end{align}
The derivative term in $q$ reflects the same mixing between the braiding and quadratic terms encountered in Sec.~\ref{subsec:closed_theory_space}.

\section{Derivation of the Minimal Ansatz}
\label{app:minimal_ansatz}
In this Appendix, we derive the minimal functional ansatz used in the main text. We start out in the more standard canonical kinetic frame, and expand about the constant-coupling cubic Galileon point and retain the lowest non-trivial field dependence of the braiding coefficient. The remaining independent quadratic operator is kept with a constant coupling coefficient.
This construction is by no means unique, but it is the simplest one-rolling-function modification of the cubic Galileon within the closed Rolling Galileon theory space.

\subsection{Construction in the Canonical Kinetic Frame}
In the canonical kinetic frame, given by \eqref{eq:app_kinframe}, the shift-symmetric cubic Galileon corresponds to the special point $\{g(\psi)=g_0, h(\psi)=0\}$. The simplest extension of this is to promote $g_0$ to a function of the field and include a constant quadratic coefficient $h_0$, $\{g(\psi), h(\psi)=h_0\}$, giving
\begin{equation}
    \mathcal L_\psi
    = -\mathcal X  
    - g(\psi) \mathcal X\Box\psi
    + h_0 \mathcal X^2.
    \label{eq:app_kinframe_ansatz}
\end{equation}
Holding $h=h_0$ fixed, rather than allowing it to roll independently, ensures that the model contains one functional freedom $g(\psi)$. 

\subsection{Specifying the Rolling Profile}
It still remains to specify that freedom. The simplest shift-symmetry breaking extension of the constant-coupling theory is a linear expansion about the cubic Galileon point $g_0$,
\begin{equation}
    g(\psi) = g_0 + g_1\psi.
  \label{eq:app_g_linear}
\end{equation}
However, we may eliminate $g_0$ through a redefinition of the field. Defining 
\begin{equation}
    \tilde \psi \coloneqq \psi + \frac{g_0}{g_1},
\end{equation}
the derivative terms remain invariant
\begin{equation}
    \partial_\mu \tilde \psi =  \partial_\mu \psi,
    \quad
    \tilde{\mathcal X} = \mathcal X, 
    \quad
    \Box \tilde \psi = \Box \psi ,
\end{equation}
where $\tilde{\mathcal X} \coloneqq -\frac12g^{\mu\nu}\partial_\mu \tilde \psi\partial_\nu \tilde \psi$. Relabelling the field $\tilde \psi\to\psi$, the action becomes
\begin{equation}
    \mathcal L_\psi
    = -\mathcal X  
    - g_1 \psi \mathcal X\Box\psi
    + h_0 \mathcal X^2,
    \label{eq:app_kinframe_min_model}
\end{equation}
where our single rolling function in this frame is
\begin{equation}
    g(\psi)=g_1\psi.
    \label{eq:simplest_rolling_function}
\end{equation}
Thus $g_0$ is a redundant coordinate offset rather than an additional free parameter.
Consequently, this action~\eqref{eq:app_kinframe_min_model} is the minimal rolling extension of the cubic Galileon and hence the simplest Rolling Galileon.



\subsection{Transformation to the Canonical Braiding Frame}
The canonical braiding frame functions are related to the canonical kinetic functions via~\eqref{eq:k_from_g} and~\eqref{eq:kinetic_braiding_inverse}. Since $h(\psi)=h_0$, the rolling profiles of $\{k,q\}$ are set by the rolling of $g(\psi)$~\eqref{eq:simplest_rolling_function} alone,
\begin{align}
    k(\psi) = \frac{1}{(g_1\psi)^{2/3}},
    \quad
    q(\psi) = \left(k_0-\frac23 g_1\right)k(\psi)^2.
    \label{eq:minimal_in_ckf}
\end{align}
We thus find that the minimal rolling model links the two canonical braiding frame functions $\{k,q\}$ via $q\propto k^2$.  

To determine the functional form of these invariants in the canonical braiding frame, we again make the transformation that normalises the braiding coefficient
\begin{equation}
    \frac{d\psi}{d\phi} = g(\psi)^{-1/3}, 
    \quad\text{or equivalently,}\quad 
    \frac{d\phi}{d\psi} = g(\psi)^{1/3}
    \label{eq:minimal_transformation}
\end{equation}
Integrating, choosing the integration constant so that $\phi=0$ at $\psi=0$, and substituting into~\eqref{eq:simplest_rolling_function} gives 
\begin{equation}
    \phi=\frac34 g_1^{1/3}\psi^{4/3}, 
    \quad\implies\quad
    g(\phi(\psi)) = \left(\frac{4g_1}{3} \phi \right)^{3/4}.
    \label{eq:minimal_g}
\end{equation}
Substituting~\eqref{eq:minimal_g} into~\eqref{eq:kinetic_braiding_inverse} yields 
\begin{equation}
  k(\phi) = \frac{k_0}{\sqrt{\phi}},
  \qquad
  q(\phi) = \frac{q_0}{\phi},
  \label{eq:app_minimal_kq}
\end{equation}
where we have defined
\begin{equation}
    k_0 \coloneqq  \sqrt{\frac{3}{4g_1}},
    \qquad
    q_0 \coloneqq \frac{3h_0 - 2g_1}{4g_1}.
\end{equation}

\begin{figure*}[t!]
    \centering

    \begin{subfigure}[t]{0.49\textwidth}
        \centering
        \includegraphics[width=\linewidth]{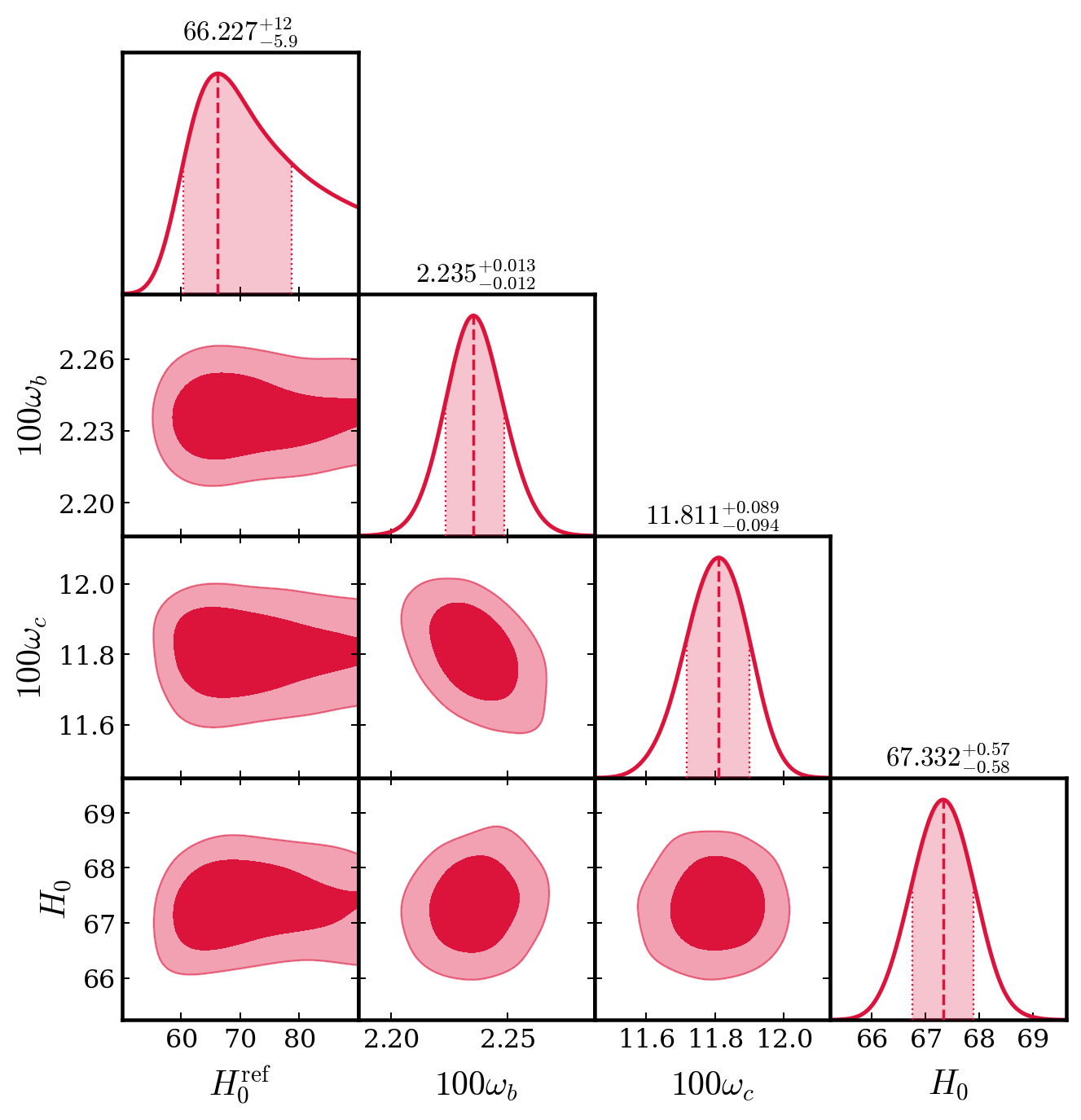}
        \caption{$k$-only model.}
        \label{fig:corner-konly_other}
    \end{subfigure}
    \hfill
    \begin{subfigure}[t]{0.49\textwidth}
        \centering
        \includegraphics[width=\linewidth]{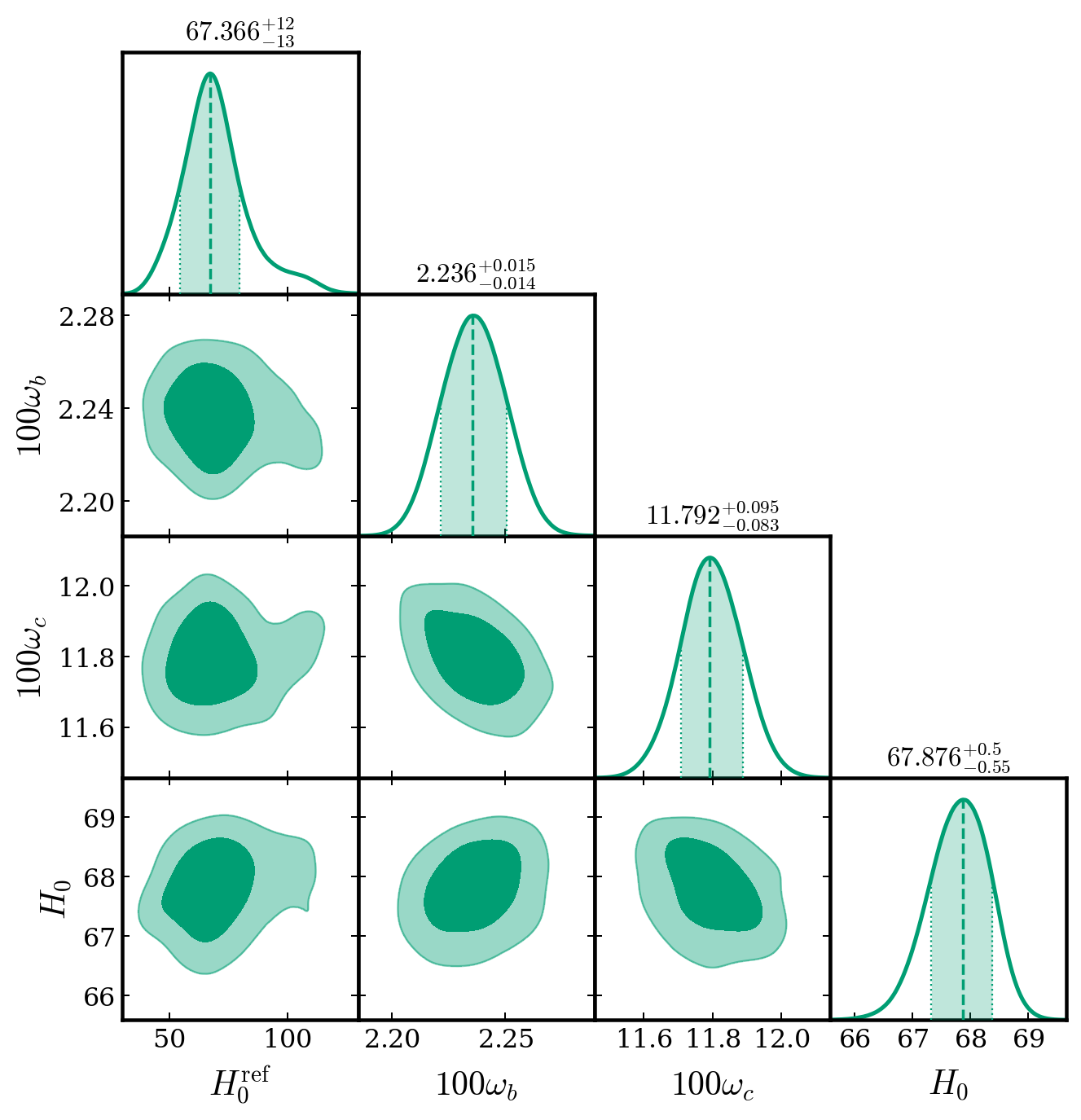}
        \caption{$k$-$q$ model.}
        \label{fig:corner-kq_other}
    \end{subfigure}

    \caption{
    Marginalised model-independent posterior constraints for the two minimal RG models fitted to the joint CMB+BAO+SN data set with the positive-ISW condition imposed. The left panel shows the $k$-only model~\eqref{eq:k-only_model}, whilst the right panel shows the full minimal $k$-$q$ model~\eqref{eq:minimal_theory}.
    }
    \label{fig:corner_other}
\end{figure*}

For $k_0,q_0>0$ and $\dot \phi>0$, both functions~\eqref{eq:app_minimal_kq} are positive and decreasing. For $q_0>0$ to be true, we must have $h_0>2g_1/3$. We cannot therefore have a vanishing quadratic strength in the canonical kinetic frame, $h_0\neq0$; it must be positive, and large enough so as to ensure $q_0>0$.  Thus the kinetic and quadratic strengths decrease relative to the canonically normalised braiding operator, precisely realising the phenomenologically preferred rolling behaviour~\eqref{eq:preferred}. The minimal Rolling Galileon action~\eqref{eq:app_kinframe_min_model} in the canonical braiding frame is therefore 
\begin{equation}
    \mathcal L_\phi
    =
    -\frac{k_0}{\sqrt{\phi}} X
    - X\Box\phi
    + \frac{q_0}{\phi} X^2.
    \label{eq:app_minimal_action}
\end{equation}



\section{Model-Independent Constraints}
\label{app:model_indep_constraints}

For completeness, we present the remaining marginalised constraints on the cosmological parameters common to both models in Fig.~\ref{fig:corner_other}. In both, the physical densities for baryons $\omega_b$ and cold dark matter $\omega_c$, and the Hubble constant $H_0$ recover their usual $\Lambda$CDM values. Recall from Sec.~\ref{sec:Hi-COLA} (see also~\cite{Wright:2022krq, Gupta:2024seu, HiCOLA:Krishna.paper}) that $H_{0}^{\mathrm{ref}}$ is used to construct the background initial conditions. It fixes the early-time normalisation and need not coincide exactly with the $H_0$ obtained after evolving the model. These posteriors therefore confirm that the conventional matter sector and early-time expansion remain essentially $\Lambda$CDM-like, with departures arising through the late-time evolution of the rolling scalar.










\section{Rolling Galileons with a Potential}
\label{app:potential}
In this Appendix, we examine the consequences of extending Rolling Galileons with a scalar potential.
We explicitly demonstrate that the inclusion of a single potential naturally induces a tension between the phantom-crossing and void-viability requirements. We note, however, that Rolling Galileons will exhibit mixed screening, whose comprehensive treatment requires solving the corresponding master equation~\cite{Sirera:2026klo}. Such analysis lies beyond the scope of this work, and our conclusions remain within the Vainshtein-only limit adopted here. 


\subsection{Theory Space}
The operator coefficients $A$, $B$ and $C$ mix under field redefinitions of the main text, and it is for this reason that the physical content is carried by the invariant pair $\{k,q\}$. A potential, however, remains invariant under $\varphi = \varphi(\phi)$,
\begin{equation}
    \widetilde V(\phi) = V(\varphi(\phi)), \footnote{If inverse powers of $X$ are admitted, a term $B_{-1}(\varphi)\mathcal X^{-1}\Box\varphi$ generates an $X^0$ contribution under nonlinear scalar-field redefinitions and can therefore mix with a potential. Such an inverse-power extension is singular at $X=0$ and lies outside the Rolling Galileon theory space considered here.}
    \label{eq:V_invariant}
\end{equation}
and so may be added unambiguously in any frame. The extended Rolling Galileon Lagrangian is then
\begin{equation}
    \mathcal L_\phi
    =
    - k(\phi) X
    - X\Box \phi
    + q(\phi) X^2
    -V(\phi),
    \label{eq:ACGV_action}
\end{equation}
which enlarges the closed two-function theory space by one further function. Of course, a constant potential is degenerate with the cosmological constant already present in the action and contributes nothing new. Only a rolling potential, $V_\phi \neq 0$, is dynamically different. 

\subsection{Phenomenology}
Because the potential carries no $X$-dependence, it leaves the braiding and the kinetic derivative structure untouched. On the FLRW background~\eqref{eq:FLRW_metric}, it shifts the scalar energy density and pressure by $\rho_\phi \to \rho_\phi + V$ and $p_\phi \to p_\phi - V$ respectively, so that their sum is unchanged,
\begin{equation}
    \rho_\phi + p_\phi
    = 2X \left( -k + 2qX + 3H\dot\phi - \ddot\phi\right).
    \label{eq:}
\end{equation}
The equation of motion is subsequently altered as 
\begin{equation}
    \ddot \phi_V
    =
    \ddot\phi-\frac{V_\phi}{\mathcal D}
    \label{eq:RGBV_eom},
\end{equation}
where $\mathcal D = 6H\dot\phi - k + 3q\dot\phi^2$ is the common denominator. Substituting~\eqref{eq:RGBV_eom} into the phantom-crossing~\eqref{eq:P_exact} and void~\eqref{eq:Q_implicit} combinations of the main text, the potential enters both in the same manner as the operator-derivative terms,
\begin{align}
    \mathcal P_V
    =\mathcal P + \frac{V_\phi}{4\mathcal D},
    \qquad 
    \mathcal F_V = \mathcal F - \frac{V_\phi}{4\mathcal D}
    \label{eq:RGBV_P}
\end{align}
The single new function thus pushes the phantom-crossing and void conditions in opposite directions, inducing the same phantom-void tension as the $k$-only model.

That is not to say the tension cannot be relieved with a potential at all. Indeed, a rolling $k(\phi)$ and a rolling $V(\phi)$, for example, can be played against each other--one carrying the crossing and the other to supply the antisymmetric rebalance that $q_\phi$ would otherwise provide. However, the potential transforms trivially under field redefinitions~\eqref{eq:V_invariant} and so takes no part in the operator mixing. It is a genuinely independent functional freedom, decoupled from the kinetic-braiding structure. The invariant $q$, by contrast, is fundamentally forced into existence by the very rolling of the braiding operator, so that a single rolling braiding function generates both $k$ and $q$ together. The minimal ansatz makes this explicit through $q(\phi) \propto k(\phi)^2$~\eqref{eq:minimal_in_ckf}, whereby one functional freedom may deliver both the crossing and its counterbalance. Resolving the tension with $\{k,V\}$ therefore demands two unlinked functions and an inherently larger parameter space, whereas $\{k,q\}$ achieves it with one. A residual cosmological constant together with a naturally linked $k$-$q$ pair is thus the absolute minimal choice within the Rolling Galileon theory space---and is precisely the one realised in the main text.


\bibliographystyle{apsrev4-2-titles}
\bibliography{ref}

\end{document}

%% file: preamble.tex
\usepackage[T1]{fontenc}
\usepackage[utf8]{inputenc}

\usepackage{microtype}      
\usepackage{amsmath,amssymb, bm}
\usepackage{graphicx}
\usepackage{mathtools}
\usepackage{accents}

\usepackage{capt-of}

\usepackage[pdftex, pdftitle={Rolling Galileons}, pdfauthor={Hallam}]{hyperref}

\hypersetup{pdftitle={},
  pdfsubject={}, 
  pdfauthor={},
  pdfkeywords={}, 
  pdfstartview=FitH,
  pdfpagemode={UseOutlines},
  bookmarksnumbered=true, bookmarksopen=true, colorlinks,
    citecolor=black,%
    filecolor=black,%
    linkcolor=black,%
    urlcolor=black
}

\usepackage{lipsum}

\usepackage{capt-of}

\makeatletter
\newenvironment{plainwidetext}
  {%
    \par
    \onecolumngrid
    \ignorespaces
  }
  {%
    \par
    \twocolumngrid
    \ignorespacesafterend
  }
\makeatother

